\documentclass[final,times,twocolumn,sort&compress]{elsarticle}
\usepackage{amsmath}
\usepackage{amssymb}
\usepackage{graphics}

\journal{Journal of Magnetism and Magnetic Materials}

\begin{document}

\begin{frontmatter}

\title{Spin frustration of a spin-1/2 Ising-Heisenberg three-leg tube as an indispensable ground for thermal entanglement\tnoteref{grant}}  
\author[upjs]{Jozef Stre\v{c}ka\corref{coraut}}
\tnotetext[grant]{This work was financially supported by Ministry of Education of Slovak Republic provided under the VEGA grant Nos.~1/0331/15 and 1/0043/16, 
by the grants of the Slovak Research and Development Agency provided under the contract Nos. APVV-0097-12 and APVV-14-0073, and by the Brazilian grant agency CAPES.
R.C. thanks CAPES for the financial support and Faculty of Science of P. J. \v{S}af\'{a}rik University in Ko\v{s}ice for the afforded hospitality.}
\cortext[coraut]{Corresponding author}
\ead{jozef.strecka@upjs.sk}
\author[ufal]{Raphael Cavalcante Al\'{e}cio} 
\author[ufal]{Marcelo L. Lyra} 
\author[ufla]{Onofre Rojas}     
\address[upjs]{Department of Theoretical Physics and Astrophysics, Faculty of Science, P. J. \v{S}af\'{a}rik University, Park Angelinum 9, 040 01 Ko\v{s}ice, Slovakia}
\address[ufal]{Instituto de F\'isica, Universidade Federal de Alagoas, 57072-970, Maceio-AL, Brazil}
\address[ufla]{Departamento de Ciencias Exatas, Universidade Federal de Lavras, 37200-000, Lavras-MG, Brazil}

\begin{abstract}
The spin-1/2 Ising-Heisenberg three-leg tube composed of the Heisenberg spin triangles mutually coupled through the Ising inter-triangle interaction is exactly solved in a zero magnetic field. By making use of the local conservation for the total spin on each Heisenberg spin triangle the model can be rigorously mapped onto a classical composite spin-chain model, which is subsequently exactly treated through the transfer-matrix method. The ground-state phase diagram, correlation functions, concurrence, Bell function, entropy and specific heat are examined in detail. It is shown that the spin frustration represents an indispensable ground for a thermal entanglement, which is quantified with the help of concurrence. The specific heat displays diverse temperature dependences, which may include a sharp low-temperature peak mimicking a temperature-driven first-order phase transition. It is convincingly evidenced that this anomalous peak originates from massive thermal excitations from the doubly degenerate ground state towards an excited state with a high macroscopic degeneracy due to chiral degrees of freedom of the Heisenberg spin triangles.
\end{abstract}


\end{frontmatter}

\section{Introduction}

Quantum spin models in one dimension traditionally attract a great deal of attention, because they often exhibit unique magnetic properties closely connected to exotic quantum ground states \cite{mattis,parkfar,diep,richter,mila}. Although all real-world magnetic materials are essentially three dimensional a lot of them can be effectively described by the notion of one-dimensional (1D) quantum Heisenberg spin models due to negligible interactions in other two spatial dimensions \cite{mila,jong74}. It should be emphasized that 1D Heisenberg spin models display more prominent quantum features than their higher-dimensional counterparts on account of reinforced quantum spin fluctuations. If the geometric spin frustration is absent, the fundamental properties of quantum Heisenberg chains basically depend on the parity of spin. The Heisenberg chains with half-odd-integer spins have a gapless excitation spectrum and algebraic decay of correlations, while the Heisenberg chains with integer spins have an energy gap and exponential decay of correlations \cite{mattis,haldane}. If the geometric spin frustration comes into play, however, the essential features of quantum Heisenberg chains may become more complex and possibilities for a low-energy spectrum are also broadened \cite{parkfar,diep,richter,mila}. 

From an immense reservoir of 1D quantum spin systems, the spin-1/2 Heisenberg tubes have recently attracted much attention \cite{cabr97,cabr98,hone00,citr00,misg03,nish08,saka08,fuji12,zhao12,arle13,gome14,yona15,kawa97,lusc04,saka05,foue06}. The term spin tube generally refers to a $n$-leg ($n \geq 3$) spin ladder with periodic boundary conditions along a rung (inter-chain) direction. The antiferromagnetic coupling along the rung direction obviously causes a geometric spin frustration whenever the odd-numbered tube is considered. Owing to this fact, the antiferromagnetic spin-$1/2$ Heisenberg three-leg tube has a spin gap in contrast to the spin-$1/2$ Heisenberg three-leg ladder with an open boundary condition along the rung (inter-chain) direction \cite{kawa97,lusc04,saka05,foue06}. The Lieb-Schultz-Mattis theorem \cite{lieb64} would suggest that the spin gap must be accompanied with at least doubly-degenerate ground state with a spontaneous breaking of the translational symmetry since the unit cell consists of three spins. From the experimental point of view, 1D coordination polymers [(CuCl$_2$tachH)$_3$Cl]Cl$_2$ (tach=1,3,5-triaminocyclohexane) \cite{seeb04,schn04,ivan10} and Cu$_2$Cl$_4$ $\cdot$ D$_8$C$_4$SO$_2$ \cite{garl08} provide unique experimental realizations of a spin-1/2 Heisenberg three-leg and four-leg tube, respectively. 

In the present work, we will exactly examine a spin frustration and thermal entanglement of the spin-1/2 Ising-Heisenberg three-leg tube, which accounts for the Heisenberg intra-triangle and Ising inter-triangle interaction. This simplified but still highly non-trivial 1D quantum spin system is exactly tractable by adapting the approach worked out previously for the spin-1/2 Ising-Heisenberg tetrahedral chain \cite{roja13,stre14}. The exotic quantum ground states along with a mutual interplay of spin frustration and quantum entanglement will be the main subject matter of our investigations. In particular, we will compare a frustration temperature \cite{sant92} with a threshold temperature of thermal entanglement, which will be calculated from a disappearance of the concurrence serving as a measure of bipartite entanglement \cite{woot98,amic04}. Besides, we will also calculate the non-locality function in order to test whether or not the Bell inequality is violated \cite{horo95}, because the nonlocality and entanglement capture different aspects of quantum correlations \cite{aule09,suny14}. 

The organization of this paper is as follows. The spin-$1/2$ Ising-Heisenberg tube will be introduced in Sec. \ref{method} along with basic steps of its exact analytical treatment. Section \ref{result} deals with the most interesting results obtained for the ground-state phase diagram, correlation functions, spin frustration, bipartite entanglement, non-locality and specific heat. Finally, several concluding remarks are mentioned in Section \ref{conclusion}.

\section{Model and method}

\label{method}
\begin{figure}
\centering
\includegraphics[scale=0.3]{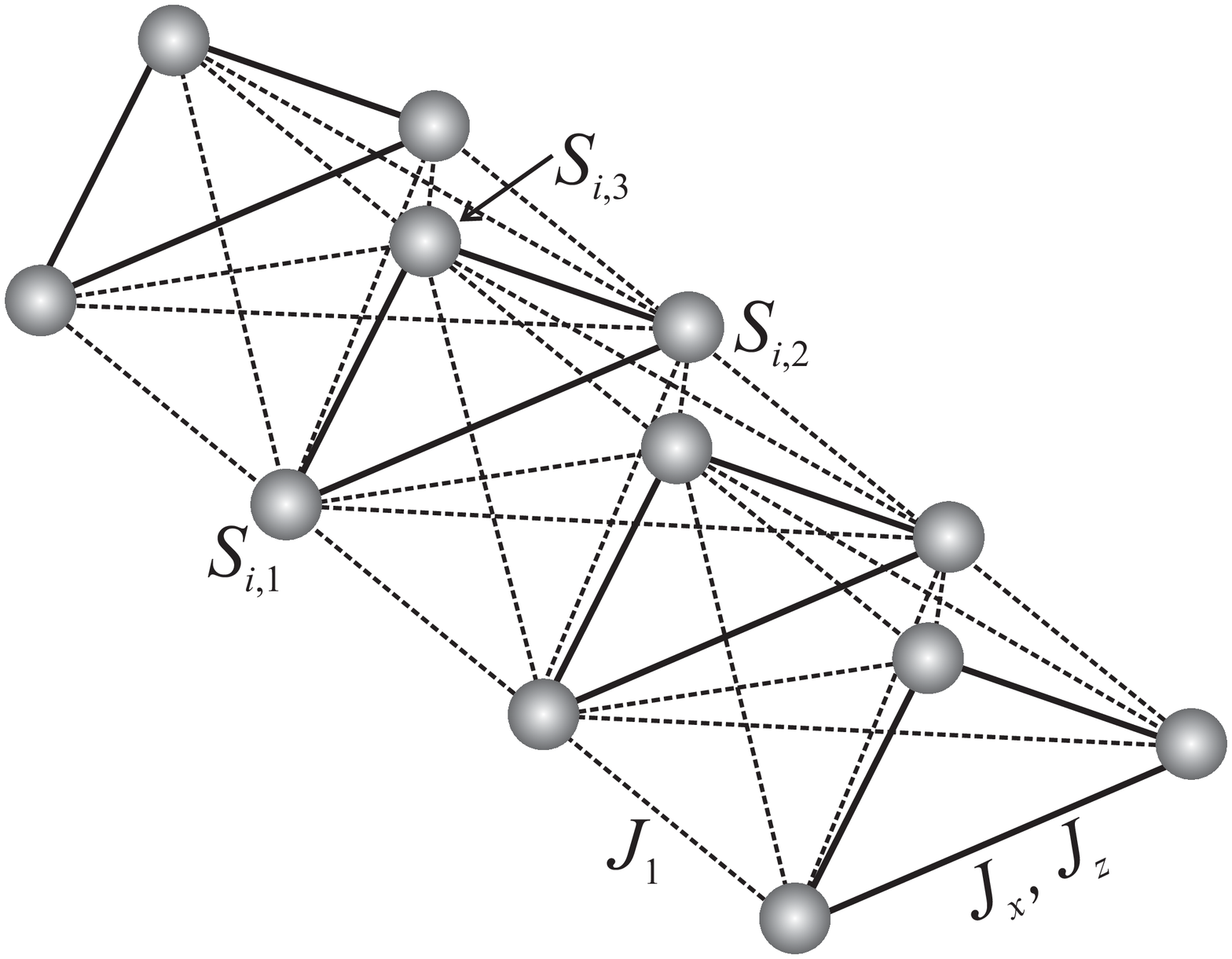}
\caption{A diagrammatic representation of spin-1/2 Ising-Heisenberg three-leg tube. Thick solid lines represent the XXZ Heisenberg intra-triangle interaction ($J_x$, $J_z$), while thin broken lines correspond to the Ising inter-triangle coupling $J_1$.}
\label{fig1}
\end{figure}

Let us consider the spin-$1/2$ Ising-Heisenberg three-leg tube with a cross-section of equilateral spin triangles, whereas the spins belonging to the same triangular unit are mutually coupled through the Heisenberg intra-triangle interaction and the spins from neighboring triangular units are coupled through the Ising inter-triangle interaction (see Fig.~\ref{fig1}). The Hamiltonian of the spin-$1/2$ Ising-Heisenberg three-leg tube is then given by
\begin{eqnarray}
\hat{H} \!\!\!&=&\!\!\! \sum_{i=1}^N\sum_{j=1}^3 \left[ J_x \left(\hat{S}_{i,j}^x \hat{S}_{i,j+1}^x + \hat{S}_{i,j}^y \hat{S}_{i,j+1}^y \right) + J_z \hat{S}_{i,j}^z \hat{S}_{i,j+1}^z \right] 
\nonumber \\
\!\!\!&+&\!\!\! J_1 \sum_{i=1}^N \left(\sum_{j=1}^3 \hat{S}_{i,j}^z \right) \left(\sum_{j=1}^3 \hat{S}_{i+1,j}^z \right),
\label{1}
\end{eqnarray}
where $\hat{S}_{i,j}^{\alpha}$ $(\alpha \in \{x,y,z\})$ mark spatial components of the standard spin-$1/2$ operator, the former subscript $i$ determines a position of a triangular unit within a spin tube and the latter subscript $j$ specifies a position of individual spin within a given triangular unit by imposing cyclic boundary conditions $\hat{S}_{i,4}^{\alpha} \equiv \hat{S}_{i,1}^{\alpha}$, $\hat{S}_{N+1,j}^{\alpha} \equiv \hat{S}_{1,j}^{\alpha}$ (see Fig.~\ref{fig1}). The interaction terms $J_x$ and $J_z$ stand for the XXZ Heisenberg intra-triangle interaction between three spins from the same triangular unit and the coupling constant $J_1$ labels the Ising inter-triangle interaction between all spins from neighboring triangular units. 

The total Hamiltonian (\ref{1}) of the spin-1/2 Ising-Heisenberg tube can be alternatively rewritten in terms of \textit{composite spin operators}, which determine the total spin of the Heisenberg triangles and its $z$-component 
\begin{eqnarray}
\hat{\bf T}_{i} = \sum_{j=1}^3 \hat{\bf S}_{i,j} \qquad \mbox{and} \qquad \hat{T}_i^z = \sum_{j=1}^3 \hat{S}_{i,j}^z.
\label{ts}
\end{eqnarray}
It can be proved by inspection that the composite spin operators $\hat{\bf T}_{i}^2$ and $\hat{T}_i^z$ commute with the total Hamiltonian (\ref{1}), i.e. $[\hat{H}, \hat{\bf T}_{i}^2] = [\hat{H}, \hat{T}_i^z] = 0$, which means that the total spin of the Heisenberg triangles and its $z$-component represent conserved quantities with well defined quantum numbers. Consequently, the eigenvalues of the total Hamiltonian (\ref{1}) can be related to the eigenvalues of the composite spin operators $\hat{\bf T}_{i}^2$ and $\hat{T}_i^z$. Using the spin identity $(\hat{T}_{i}^{\alpha})^2 = \frac{3}{4} - 2(\hat{S}_{i,1}^{\alpha} \hat{S}_{i,2}^{\alpha} + \hat{S}_{i,2}^{\alpha} \hat{S}_{i,3}^{\alpha} + \hat{S}_{i,3}^{\alpha} \hat{S}_{i,1}^{\alpha})$, the total Hamiltonian (\ref{1}) can be rewritten into the following form
\begin{equation}
\hat{H} = - \frac{3N}{8}(2J_x + J_z) + \sum_{i=1}^N \hat{H}_i,
\label{2}
\end{equation}
which depends on the Hamiltonian $\hat{H}_i$ of two subsequent triangular unit cells 
\begin{eqnarray}
\hat{H}_i \!\!\!&=&\!\!\! J_1 \hat{T}_i^z \hat{T}_{i+1}^z + \frac{J_x}{4} (\hat{\bf T}_{i}^2 + \hat{\bf T}_{i+1}^2) \nonumber \\
\!\!\!&+&\!\!\! \frac{J_z - J_x}{4}[(\hat{T}_i^z)^2 + (\hat{T}_{i+1}^z)^2].
\label{3}
\end{eqnarray}
The first term in Eq. (\ref{2}) is the less important constant term and the second one is expressed as a sum over the symmetrically defined cell Hamiltonians $\hat{H}_i$, which depend according to Eq. (\ref{3}) on the composite spin operators from two neighboring triangular units. The eigenvalues of the composite spin operators $\hat{\bf T}_{i}^2$ and $\hat{T}_i^z$ are quantized according to the rules $T_i(T_i + 1)$ with $T_i = 1/2$ or $3/2$ and, respectively, $T_i^z = -T_i,-T_i+1, \dots,T_i$. From this point of view, the spin-1/2 Ising-Heisenberg tube defined by the Hamiltonian (\ref{1}) can be rigorously mapped onto some classical composite spin-chain model, which can be further treated by the transfer-matrix method \cite{baxter}. Owing to a validity of the commutation relation between the cell Hamiltonians $[\hat{H}_i , \hat{H}_j] = 0$, the partition function can be factorized into the following form 
\begin{eqnarray}
Z \!\!\!&=&\!\!\! \exp \left[ \frac{3N\beta}{8} (2J_x + J_z) \right] \mbox{Tr}  \prod_{i=1}^N   \exp(-\beta \hat{H}_i) \nonumber \\
\!\!\!&=&\!\!\! \exp\left[ \frac{3N\beta}{8}(2J_x + J_z )\right] \sum_{\{T_i, T_i^z\}} \prod_{i=1}^N  \mbox{W}(T_i, T_i^z;T_{i+1}, T_{i+1}^z) \nonumber \\
\!\!\!&=&\!\!\! \exp\left[ \frac{3N\beta}{8} (2J_x  + J_z) \right] \mbox{Tr} \, \mbox{W}^N,
\label{tm}
\end{eqnarray}
where $\beta = 1/(k_{\rm B} T)$, $k_{\rm B}$ is the Boltzmann's constant, $T$ is the absolute temperature, and the summation $ \sum_{\{T_i, T_i^z\}}$ runs over all possible values of the quantum spin numbers $T_i$ and $T_i^z$. The expression $W = \exp(-\beta \hat{H}_i)$, which depends on the composite spin operators from two neighboring triangular units, can be alternatively viewed as the transfer matrix with the following elements
{\tiny  \begin{eqnarray}
\!\!\!\!\!\!\!\!\!&&\!\!\!\!\!\!\!\!\! W (T_i, T_i^z; T_{i+1}, T_{i+1}^z) = \langle T_i, T_i^z | {\rm e}^{-\beta \hat{H}_i} |T_{i+1}, T_{i+1}^z \rangle = \label{tmm} \\
\!\!\!\!\!\!\!\!\!&&\!\!\!\!\!\!\!\!\! \left(\!\!\begin{array}{cccccccc}
x^3 y^9 z^9 & x^5 y^5 z^3 & x^5 y^5 z^{-3} & x^3 y^9 z^{-9} & x^2 y^5 z^3 & x^2 y^5 z^{-3} & x^2 y^5 z^3 & x^2 y^5 z^{-3} \\
x^5 y^5 z^3 & x^7 y z & x^7 y z^{-1} & x^5 y^5 z^{-3} & x^4 y z & x^4 y z^{-1} & x^4 y z & x^4 y z^{-1} \\
x^5 y^5 z^{-3} & x^7 y z^{-1} & x^7 y z & x^5 y^5 z^{3} & x^4 y z^{-1} & x^4 y z & x^4 y z^{-1} & x^4 y z \\
x^3 y^9 z^{-9} & x^5 y^5 z^{-3} & x^5 y^5 z^3 & x^3 y^9 z^{9} & x^2 y^5 z^{-3} & x^2 y^5 z^3 & x^2 y^5 z^{-3} & x^2 y^5 z^3  \\
x^2 y^5 z^{3} & x^4 y z & x^4 y z^{-1} & x^2 y^5 z^{-3} & x y z & x y z^{-1} & x y z & x y z^{-1}  \\
x^2 y^5 z^{-3} & x^4 y z^{-1} & x^4 y z & x^2 y^5 z^{3} & x y z^{-1} & x y z & x y z^{-1} & x y z  \\
x^2 y^5 z^{3} & x^4 y z & x^4 y z^{-1} & x^2 y^5 z^{-3} & x y z & x y z^{-1} & x y z & x y z^{-1}  \\
x^2 y^5 z^{-3} & x^4 y z^{-1} & x^4 y z & x^2 y^5 z^{3} & x y z^{-1} & x y z & x y z^{-1} & x y z \nonumber
\end{array}\!\! \right),
\end{eqnarray}}
where $x =  \exp(-\beta J_x/4)$, $y =  \exp(-\beta J_z/8)$ and $z = \exp(-\beta J_1/4)$. As usual, the partition function is in the thermodynamic limit $N \to \infty$ solely determined by the largest eigenvalue of the transfer matrix $W$ given by Eq. (\ref{tmm}). By inspection, four out of eight transfer-matrix eigenvalues equal zero ($\lambda_5 = \lambda_6 = \lambda_7 = \lambda_8 = 0$), because the second, fifth and seventh (third, sixth and eigth) rows are linearly dependent. The other four eigenvalues can be found by solving two quadratic equations 
\begin{equation}
(\lambda^2 - a\lambda + b)(\lambda^2 - c\lambda + d) =0
\label{qe}
\end{equation}
with the coefficients $a$, $b$, $c$ and $d$ given by
\begin{eqnarray}
a \!\!\!&=&\!\!\! xy[x^2y^8(z^9 + z^{-9}) + (z+z^{-1})(2+x^6)], \nonumber \\
b \!\!\!&=&\!\!\! x^4 y^{10}(2+x^6)[ (z + z^{-1})(z^9 + z^{-9}) -(z^3 + z^{-3})^2], \nonumber \\
c \!\!\!&=&\!\!\! xy[x^2y^8(z^9 - z^{-9}) + (z-z^{-1})(2+x^6)], \nonumber \\
d \!\!\!&=&\!\!\! x^4 y^{10}(2+x^6)[ (z - z^{-1})(z^9 - z^{-9}) -(z^3 - z^{-3})^2]. 
\label{coe1}
\end{eqnarray}
The remaining four eigenvalues of the transfer matrix (\ref{tmm}) can be therefore acquired by solving two quadratic equations (\ref{qe})
\begin{eqnarray}
\lambda_{\pm} \!\!\!&=&\!\!\! x^3 y^9 \cosh\left(\frac{9\beta J_1}{4}\right) + x y(2+x^6) \cosh\left(\frac{\beta J_1}{4}\right) \pm \sqrt{D_1}, \nonumber \\
\lambda_{3,4} \!\!\!&=&\!\!\! x^3 y^9 \sinh\left(\frac{9\beta J_1}{4}\right) + x y(2+x^6) \sinh\left(\frac{\beta J_1}{4}\right) \pm \sqrt{D_2}, \nonumber \\
\label{eig}
\end{eqnarray}
where
\begin{eqnarray}
D_1 \!\!\!&=&\!\!\! \left[x^3 y^9 \cosh\left(\frac{9\beta  J_1}{4}\right) - x y(2+x^6) \cosh\left(\frac{\beta  J_1}{4}\right)\right]^2 \nonumber \\
    \!\!\!&+&\!\!\! 4 x^4 y^{10} (2+x^6) \cosh^2\left(\frac{3\beta J_1}{4}\right), \nonumber \\
D_2 \!\!\!&=&\!\!\! \left[x^3 y^9 \sinh\left(\frac{9\beta  J_1}{4}\right) - x y(2+x^6) \sinh\left(\frac{\beta  J_1}{4}\right)\right]^2  \nonumber \\
    \!\!\!&+&\!\!\! 4 x^4 y^{10} (2+x^6) \sinh^2\left(\frac{3\beta J_1}{4}\right).
\label{dis}
\end{eqnarray}
In thermodynamic limit $N \to \infty $, the Helmholtz free energy per unit cell is determined just by the largest transfer-matrix eigenvalue $\lambda_{\rm max} = \lambda_{+}$ 
\begin{equation}
f =  - \beta^{-1} \lim_{ N \rightarrow \infty} \frac{1}{N} \ln  \mathrm{Z} = - \frac{3(2J_x + J_z)}{8} - \beta^{-1} \ln \lambda_{+}.
\label{12}
\end{equation}
After substituting the largest transfer-matrix eigenvalue (\ref{eig}) into Eq. (\ref{12}) and straightforward algebraic manipulations one obtains an explicit form of the Helmholtz free energy 
\begin{equation}
f =  - \beta^{-1} \ln \left(g_1 + g_2 + g_4 \right),
\label{free}
\end{equation}
which is expressed in terms of the newly defined functions $g_1$--$g_4$ given by
\begin{eqnarray}
g_1 \!\!\!&=& \!\!\! \exp \left(- \frac{3 \beta J_z}{4} \right) \cosh \left(\frac{9 \beta J_1}{4} \right), \nonumber \\
g_2 \!\!\!&=& \!\!\! \exp \left[\frac{\beta}{4} \left(J_z + 2 J_x \right) \right] \left[2 + \exp\left(-\frac{3 \beta J_x}{2} \right) \right] \cosh \left(\frac{\beta J_1}{4} \right), \nonumber \\
g_3 \!\!\!&=& \!\!\! \exp \left[\frac{\beta}{2} \left(J_x - J_z \right) \right] \left[2 + \exp\left(-\frac{3 \beta J_x}{2} \right) \right] \cosh^2 \left(\frac{3\beta J_1}{4} \right), \nonumber \\
g_4 \!\!\!&=& \!\!\! \sqrt{(g_1-g_2)^2 + 4g_3}. \nonumber \\
\label{coefcorfree}
\end{eqnarray}

Now, let us calculate pair correlation functions between two spins from the same Heisenberg triangle. Both different spatial components of the pair correlation function between two spins from the same Heisenberg triangle can be calculated by differentiating the Helmholtz free energy (\ref{free}) with respect to the coupling constant $J_z$ or $J_x$. This procedure yields for the respective spatial components of the pair correlation function the following simple expressions
\begin{eqnarray}
C_{11}^{zz} \!\!\!&=&\!\!\! \langle \hat{S}_{j,i}^z \hat{S}_{j+1,i}^z \rangle = \frac{1}{Z} \mbox{Tr} \, [\hat{S}_{j,i}^z \hat{S}_{j+1,i}^z  \exp(-\beta \hat{H})] \nonumber \\
            \!\!\!&=&\!\!\! \frac{1}{4} \frac{\left(g_1 - g_2 \right) \left(g_1 + \frac{g_2}{3} \right) + \frac{4 g_3}{3} + \left(g_1 - \frac{g_2}{3}\right) g_4}
                                        {\left(g_1 - g_2 \right)^2 + 4 g_3 + \left(g_1 + g_2 \right) g_4}, \label{czz} \\
C_{11}^{xx} \!\!\!&=&\!\!\! \langle \hat{S}_{j,i}^x \hat{S}_{j+1,i}^x \rangle  = \langle \hat{S}_{j,i}^y \hat{S}_{j+1,i}^y \rangle \nonumber \\
            \!\!\!&=&\!\!\! \frac{1}{Z}  \mbox{Tr} \, [\hat{S}_{j,i}^x \hat{S}_{j+1,i}^x \exp(-\beta \hat{H}) ] \nonumber \\
            \!\!\!&=&\!\!\! \frac{1}{6} \frac{g_5 \left(g_1 - g_2 \right) - 2 g_6 - g_4 g_5}
                                        {\left(g_1 - g_2 \right)^2 + 4 g_3 + \left(g_1 + g_2 \right) g_4}, 
\label{cxx}
\end{eqnarray}
which contain two new functions $g_5$--$g_6$ given by
\begin{eqnarray}
g_5 \!\!\!&=& \!\!\! \exp \left[\frac{\beta}{4} \left(J_z + 2 J_x \right) \right] \left[1 - \exp\left(-\frac{3 \beta J_x}{2} \right) \right] \cosh \left(\frac{\beta J_1}{4} \right), \nonumber \\
g_6 \!\!\!&=& \!\!\! \exp \left[\frac{\beta}{2} \left(J_x - J_z \right) \right] \left[1 - \exp\left(-\frac{3 \beta J_x}{2} \right) \right] \cosh^2 \left(\frac{3\beta J_1}{4} \right). \nonumber \\
\label{coefcor}
\end{eqnarray}
The $z$-component of the pair correlation function between two spins from the neighboring Heisenberg triangles can in turn be calculated by differentiating the Helmholtz free energy (\ref{12}) with respect to the coupling constant $J_1$ 
\begin{eqnarray}
C_{12}^{zz} \!\!\!&=&\!\!\! \langle \hat{S}_{j,i}^z \hat{S}_{j,i+1}^z \rangle = \frac{1}{Z}  \mbox{Tr} \, [\hat{S}_{j,i}^z \hat{S}_{j,i+1}^z \exp(-\beta \hat{H})] \nonumber \\
 \!\!\!&=&\!\!\! -\frac{1}{4} \frac{\left(g_1 - g_2 \right) \left(g_7 + \frac{g_8}{3} \right) + \frac{g_9}{3} + \left(g_7 + \frac{g_8}{9} \right) g_4}
                                   {\left(g_1 - g_2 \right)^2 + 4 g_3 + \left(g_1 + g_2 \right) g_4}, 
\label{c12} 
\end{eqnarray}
whereas the functions $g_7$--$g_9$ are defined as
\begin{eqnarray}
g_7 \!\!\!&=& \!\!\! \exp \left(- \frac{3 \beta J_z}{4} \right) \sinh \left(\frac{9 \beta J_1}{4} \right), \nonumber \\
g_8 \!\!\!&=& \!\!\! \exp \left[\frac{\beta}{4} \left(J_z + 2 J_x \right) \right] \left[2 + \exp\left(-\frac{3 \beta J_x}{2} \right) \right] \sinh \left(\frac{\beta J_1}{4} \right), \nonumber \\
g_9 \!\!\!&=& \!\!\! \exp \left[\frac{\beta}{2} \left(J_x - J_z \right) \right] \left[2 + \exp\left(-\frac{3 \beta J_x}{2} \right) \right] \sinh \left(\frac{3\beta J_1}{2} \right). \nonumber \\
\label{coefcor12}
\end{eqnarray}

At this stage, one may employ two spatial components of the pair correlation function (\ref{czz}) and (\ref{cxx}) between the spins from the same Heisenberg triangle in order to calculate the concurrence \cite{amic04}
\begin{equation}
\mathrm{C} =  \mbox{max} \left \{ 0 ; 4|C_{xx}| - \left|\frac{1}{2} + 2C_{zz} \right| \right \}
\label{conc}
\end{equation}
and the Bell function \cite{horo95}
\begin{equation}
\mathrm{B} = 8 \, \mbox{max} \left \{ \sqrt{C_{xx}^2 + C_{zz}^2} ; \sqrt{2 C_{xx}^2} \right \},
\label{bell}
\end{equation}
which may serve as a measure of bipartite entanglement and quantum non-locality at zero as well as non-zero temperature. The Heisenberg spins from the same triangular unit display a thermal entanglement just if the concurrence is greater than zero, i.e. $\mathrm{C}>0$, otherwise they become disentangled. On the other hand, the violation of the Bell inequality $\mathrm{B} \le 2$ can be used in order to prove a non-local character of quantum correlations. The obtained exact results for the thermal entanglement will be confronted with the ones for non-locality, because the entanglement and non-locality capture closely related yet independent features of quantum correlations \cite{horo95,aule09}. 

\section{Results and discussion}
\label{result}

Let us proceed to a discussion of the  most interesting results for the spin-1/2 Ising-Heisenberg three-leg tube by considering the particular case with the antiferromagnetic inter-triangle interaction $J_1 > 0$, which will henceforth serve as an energy unit $J_1 = 1$ (the Boltzmann's constant is also set to unity $k_{\rm B} = 1$ for easy notation). It should be mentioned that another particular case with the ferromagnetic inter-triangle interaction $J_1 < 0$ behaves quite analogously, because the spin states on each second Heisenberg triangle are merely inverted under the transformation 
$J_1 \to -J_1$. 

\subsection{Ground state}

A diagonal form of the Hamiltonian (\ref{3}) can be straightforwardly used in order to obtain all possible ground states, since the lowest-energy eigenstate of the cell Hamiltonian (\ref{3}) can be readily extended to the whole spin-1/2 Ising-Heisenberg three-leg tube. Consequently, the ground state of the spin-1/2 Ising-Heisenberg tube can be written as a tensor product over the lowest-energy eigenstate of the cell Hamiltonian (\ref{3}). One finds by inspection just three different ground states, namely, the classical antiferromagnetic phase (CAF)
\begin{equation}
|\mbox{CAF} \rangle = \prod_{j=1}^{N/2} |\! \uparrow \uparrow \uparrow \rangle_{2j-1} \otimes  |\! \downarrow \downarrow \downarrow \rangle_{2j}, 
\label{caf}
\end{equation}
the symmetric quantum trimerized phase (SQT) 
\begin{eqnarray}
|\mbox{SQT} \rangle \!\!\!&=&\!\!\! \prod_{j=1}^{N/2} \frac{1}{\sqrt{3}} {\big(}|\! \downarrow \uparrow \uparrow \rangle + |\! \uparrow \downarrow  \uparrow \rangle + |\! \uparrow \uparrow \downarrow \rangle {\big)_{2j-1}} 
\nonumber \\
\!\!\!&\otimes&\!\!\! \frac{1}{\sqrt{3}} {\big(}|\! \downarrow \downarrow \uparrow \rangle + |\! \uparrow \downarrow  \downarrow \rangle + |\! \downarrow \uparrow \downarrow \rangle {\big)_{2j}},
\label{qaf}
\end{eqnarray}
and the macroscopically degenerate chiral antiferromagnetic phase (DCA)
\begin{eqnarray} 
|\mbox{DCA} \rangle \!\!\!&=&\!\!\!  \prod_{j=1}^{N/2} 
 \left\{\begin{array}{cc} \!\!
\frac{1}{\sqrt{3}} {\big(}|\! \downarrow \uparrow \uparrow \rangle + {\rm e}^{{\rm i} \frac{2 \pi}{3}} |\! \uparrow \downarrow  \uparrow \rangle +  {\rm e}^{{\rm i} \frac{4 \pi}{3}}  |\! \uparrow \uparrow \downarrow \rangle \big)_{2j-1}   \\ \!\!
\frac{1}{\sqrt{3}} {\big(}|\! \downarrow \uparrow \uparrow \rangle + {\rm e}^{{\rm i} \frac{4 \pi}{3}}  |\! \uparrow \downarrow  \uparrow \rangle + {\rm e}^{{\rm i} \frac{2 \pi}{3}}  |\! \uparrow \uparrow \downarrow \rangle \big)_{2j-1} 
\end{array} \right. \nonumber \\
\!\!\!&\otimes&\!\!\!  
\left\{\begin{array}{cc} \!\!
\frac{1}{\sqrt{3}} {\big(}|\! \downarrow \downarrow \uparrow \rangle + {\rm e}^{{\rm i} \frac{2 \pi}{3}}  |\! \downarrow \uparrow \downarrow \rangle +  {\rm e}^{{\rm i} \frac{4 \pi}{3}}  
|\! \uparrow \downarrow \downarrow \rangle \big)_{2j} \\ \!\!
\frac{1}{\sqrt{3}} {\big(}|\! \downarrow \downarrow \uparrow \rangle + {\rm e}^{{\rm i} \frac{4 \pi}{3}}  |\! \downarrow \uparrow \downarrow \rangle +  {\rm e}^{{\rm i} \frac{2 \pi}{3}}  
|\! \uparrow \downarrow \downarrow \rangle \big)_{2j}
\end{array} \right.\!\!.
\label{dcp}
\end{eqnarray}
Owing to the time-reversal symmetry, the alternative representation of the ground states CAF, SQT and DCA can be obtained from Eqs. (\ref{caf})--(\ref{dcp}) by inter-changing the eigenkets on odd and even Heisenberg triangles, respectively. Hence, it follows that the ground states CAF and SQT are two-fold degenerate in contrast to the ground state DCA, which is $2 \times 2^{N}$ degenerate due to the time-reversal symmetry and two chiral degrees of freedom on each Heisenberg spin triangle. The spin arrangement inherent to the three available ground states is consistent with the following asymptotic values of the pair correlation functions as calculated from Eqs. (\ref{czz}), (\ref{cxx}) and (\ref{c12}) in a zero-temperature limit
\begin{eqnarray}
\mbox{CAF:} \,\, C_{11}^{xx} \!\!\!&=& \!\!\! 0, \, \qquad C_{11}^{zz} = \frac{1}{4}, \qquad C_{12}^{zz} = - \frac{1}{4}; \nonumber \\
\mbox{SQT:} \,\, C_{11}^{xx} \!\!\!&=& \!\!\! \frac{1}{6}, \qquad C_{11}^{zz} = - \frac{1}{12}, \! \quad C_{12}^{zz} = - \frac{1}{36}; \nonumber \\
\mbox{DCA:} \,\, C_{11}^{xx} \!\!\!&=& \!\!\! - \frac{1}{12}, \quad C_{11}^{zz} = - \frac{1}{12}, \! \quad C_{12}^{zz} = - \frac{1}{36}.
\label{gscor}
\end{eqnarray}  

\begin{figure}[!thb]
\centering
\includegraphics[scale=0.25]{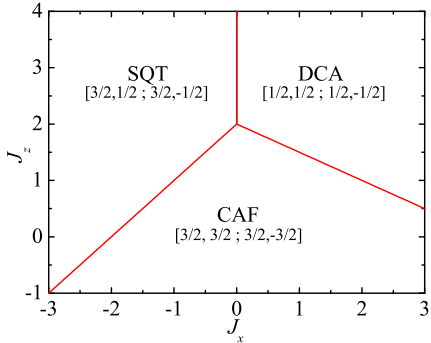}
\vspace{-0.5cm}
\caption{The ground-state phase diagram of the spin-1/2 Ising-Heisenberg three-leg tube in the $J_x-J_z$ plane. The numbers quoted in square brackets determine the total spin and its $z$-component on two consecutive Heisenberg triangles $[T_{2i-1}, T_{2i-1}^z; T_{2i}, T_{2i}^z]$.}
\label{fig2}
\end{figure}

The ground-state phase diagram involving all three available ground states is depicted in Fig. \ref{fig2}. The CAF phase becomes the ground state in a parameter space delimited by the conditions $J_z < 2 + J_x$ and $J_z < 2 - J_x/2$, which are consistent with the ferromagnetic Heisenberg interaction ($J_z<0$) or the sufficiently weak antiferromagnetic Heisenberg interaction ($J_z<2$). If the conditions $J_x > 0 $ and $J_z > 2 - J_x/2$ are met, however, the spin frustration arising out from the stronger antiferromagnetic Heisenberg interaction gives rise to the macroscopically degenerate DCA ground state with two (right- or left-hand-side) chiral degrees of freedom  on each Heisenberg triangle. As long as the conditions $J_x < 0 $ and $J_z > 2 + J_x$ are fulfilled, the SQT phase with a regular alternation of the symmetric quantum superposition of three up-up-down and down-down-up states on odd and even triangles (or vice versa) becomes the relevant ground state.

\subsection{Correlation functions}

\begin{figure}[!thb]
\centering
\includegraphics[scale=0.25]{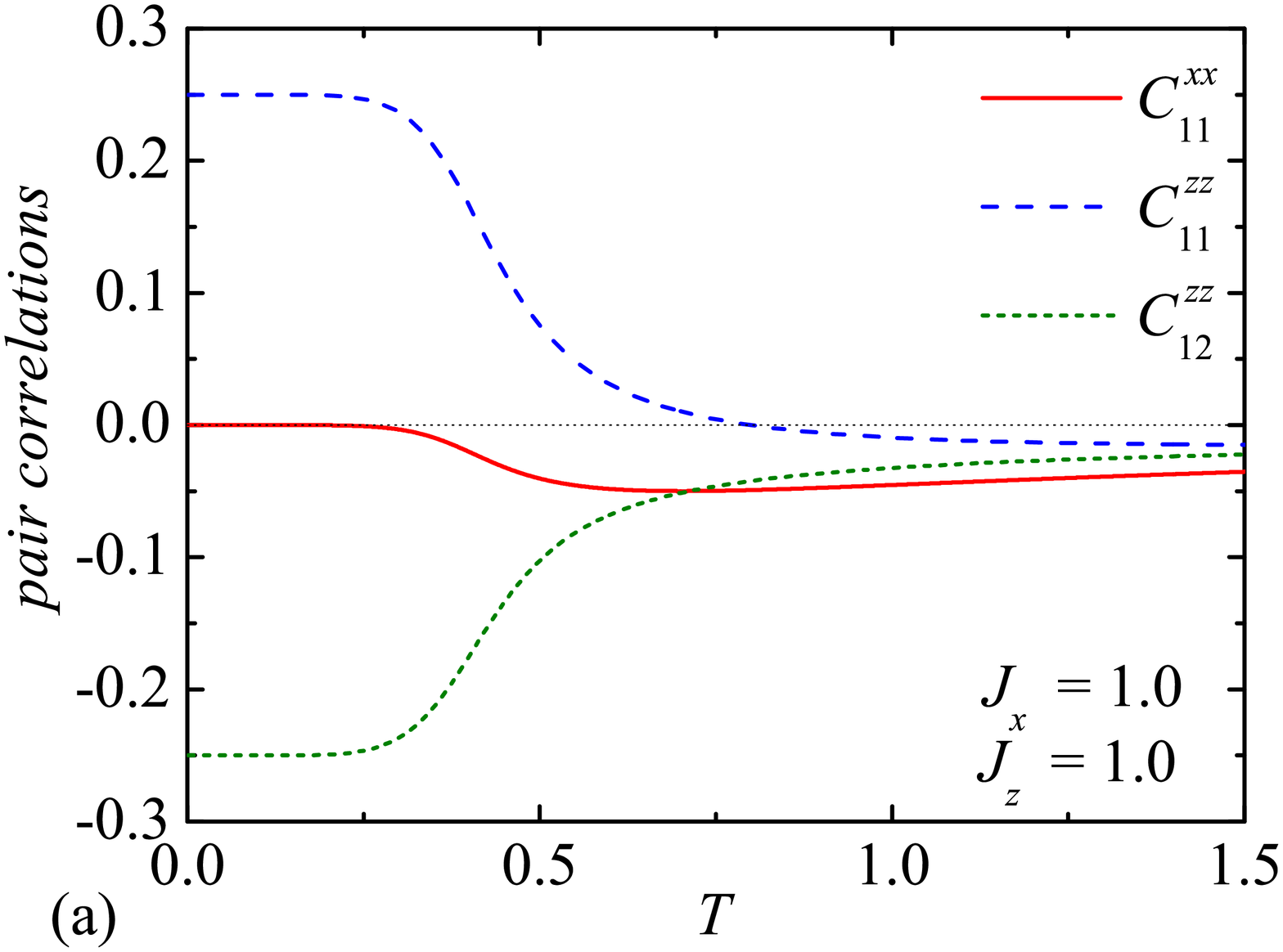}
\includegraphics[scale=0.25]{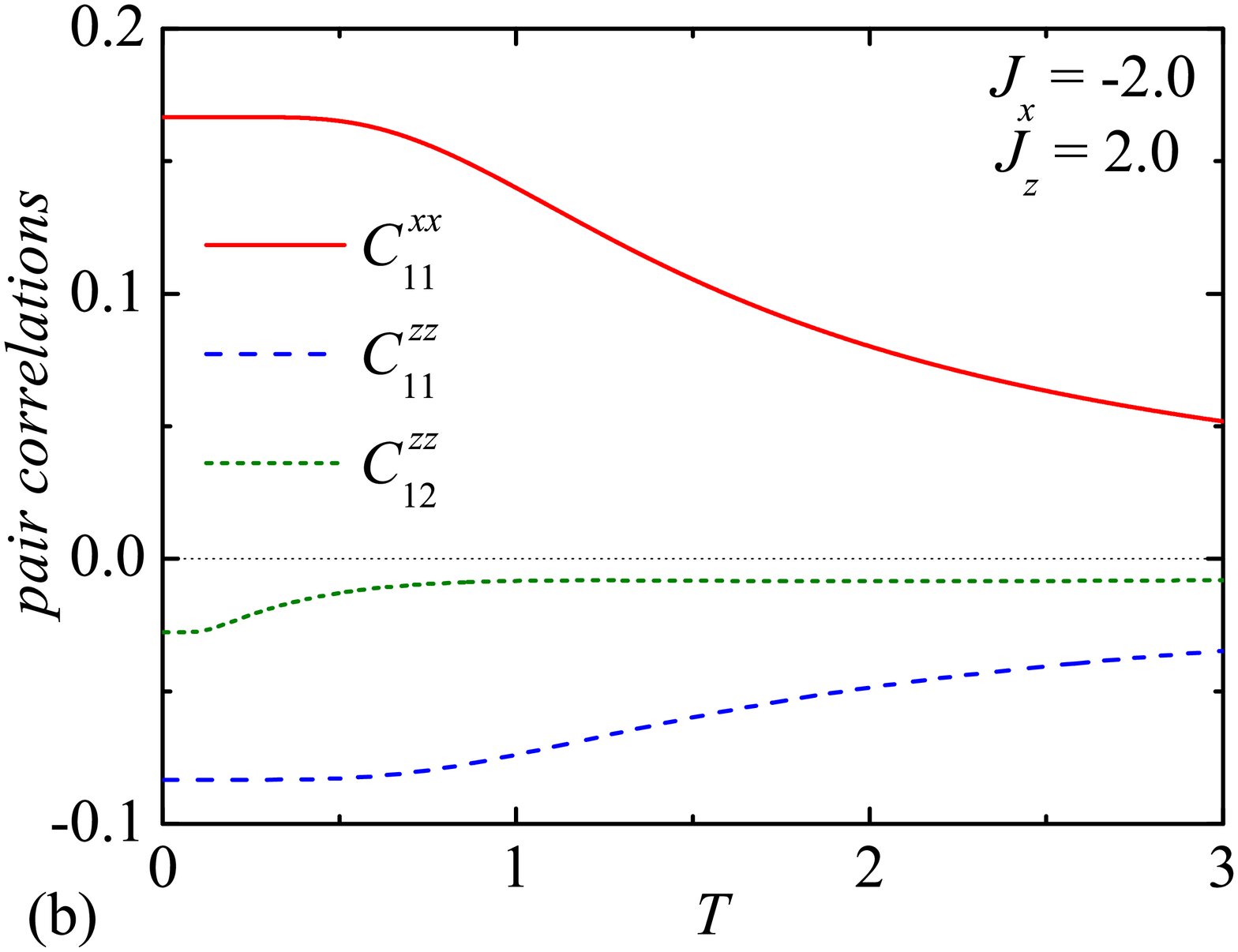}
\includegraphics[scale=0.25]{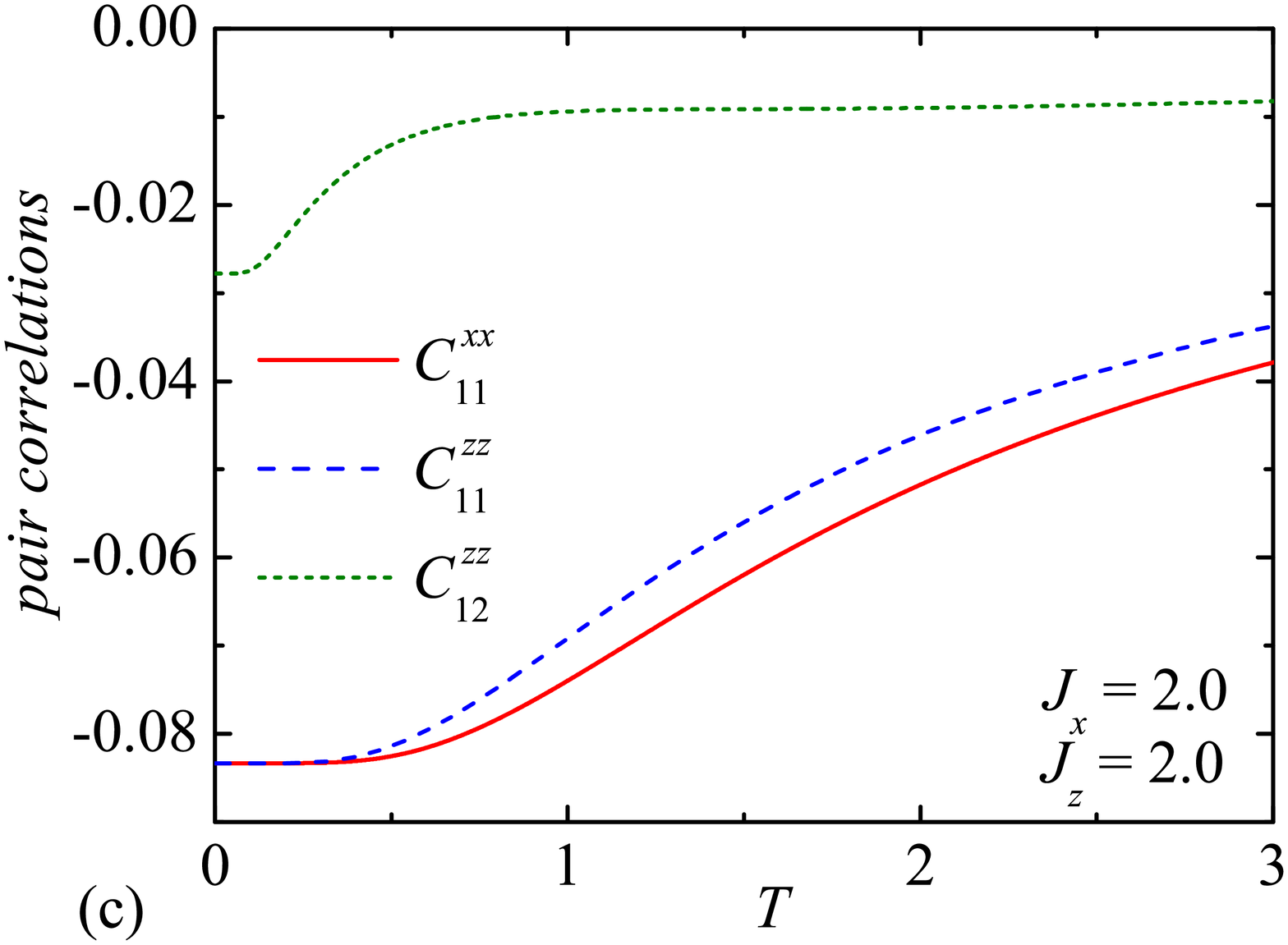}
\vspace{-0.5cm}
\caption{The pair correlation functions as a function of temperature for the coupling constants supporting three different ground states: 
(a) $J_x = 1$, $J_z = 1$ (CAF phase); (b) $J_x = -2$, $J_z = 2$ (SQT phase); (c) $J_x = 2$, $J_z = 2$ (DCA phase).}
\label{fig3}
\end{figure}

To gain an overall insight into a character of spin arrangements emerging within the individual ground states, let us explore in detail temperature dependences of all calculated pair correlation functions. The pair correlation functions are plotted against temperature in Fig. \ref{fig3}(a)-(c) for three different sets of the interaction parameters, which drive the investigated system towards the CAF, SQT and DCA ground states, respectively. It is quite clear from Fig. \ref{fig3}(a) that $z$-components of the spins from the neighboring Heisenberg triangles are perfectly anticorrelated at zero temperature, whereas a relative strength of the antiferromagnetic correlation gradually decreases with increasing temperature. Interestingly, the longitudinal correlation function between the spins from the same Heisenberg triangle shows a peculiar crossover at a so-called frustration temperature $T_f$, at which $z$-components of the spins become completely uncorrelated (i.e. the relevant correlation function equals zero). The longitudinal correlation between the spins from the same Heisenberg triangle would suggest that the $z$-components of the spins are ferromagnetically correlated below 
the frustration temperature ($T<T_f$) and antiferromagnetically correlated above it ($T>T_f$). The transverse correlation function between the spins from the same Heisenberg triangle is zero at absolute zero temperature due to a classical character of the CAF ground state, but it implies antiferromagnetic (ferromagnetic) correlation at non-zero temperatures provided that the $x$-component of the Heisenberg coupling is antiferromagnetic (ferromagnetic).

Fig. \ref{fig3}(b) demonstrates thermal variations of the correlation functions, which are quite typical for the SQT ground state. The correlation function between the spins from the same Heisenberg triangle serves in evidence of the antiferromagnetic (ferromagnetic) correlation in a longitudinal (transverse) direction, whereas a relative strength of the ferromagnetic transverse correlation is slightly stronger than that of the antiferromagnetic longitudinal correlation. Furthermore, the $z$-components of the spins from the neighboring Heisenberg triangles are antiferromagnetically correlated within the SQT phase. 

Last but not least, the correlation functions plotted in Fig. \ref{fig3}(c) illustrate typical temperature dependences for the DCA ground state. As one can see, the longitudinal and transverse correlation functions between the spins from the same Heisenberg triangle are antiferromagnetic. While the longitudinal and transverse correlation are of equal strength at zero temperature, the transverse correlation overwhelms over the longitudinal one at non-zero temperatures. It is noteworthy that the $z$-components of the spins from the neighboring Heisenberg triangles are always antiferromagnetically correlated when the investigated spin system starts from the DCA ground state. 

\subsection{Spin frustration}

It is obvious from previous discussions that the SQT and DCA ground states have frustrated character in contrast to the unfrustrated CAF ground state. According to the frustration concept developed by Toulouse \cite{toul77}, the spin system is said to be geometrically frustrated if a product of signs of the exchange couplings along an elementary plaquette becomes negative. Analogously, the product of signs of the pair correlation functions along an elementary plaquette can be used as another useful criterion for testing whether or not a spin system is frustrated at finite temperatures \cite{sant92}. Hence, the antiferromagnetic (negative) correlation function between $z$-components of the spins from the same Heisenberg triangle indeed verifies the frustrated character of the SQT and DCA phases at non-zero temperatures [see Fig. \ref{fig3}(b)-(c)]. On the other hand, the longitudinal correlation function between the spins from the same Heisenberg triangle shown in Fig. \ref{fig3}(a) changes its sign from positive (at lower temperatures) to negative (at higher temperatures), which confirms an outstanding thermally activated spin frustration above the unfrustrated CAF ground state on assumption that the antiferromagnetic intra-triangle coupling $J_z>0$ is considered. 

\begin{figure}[!thb]
\centering
\includegraphics[scale=0.25]{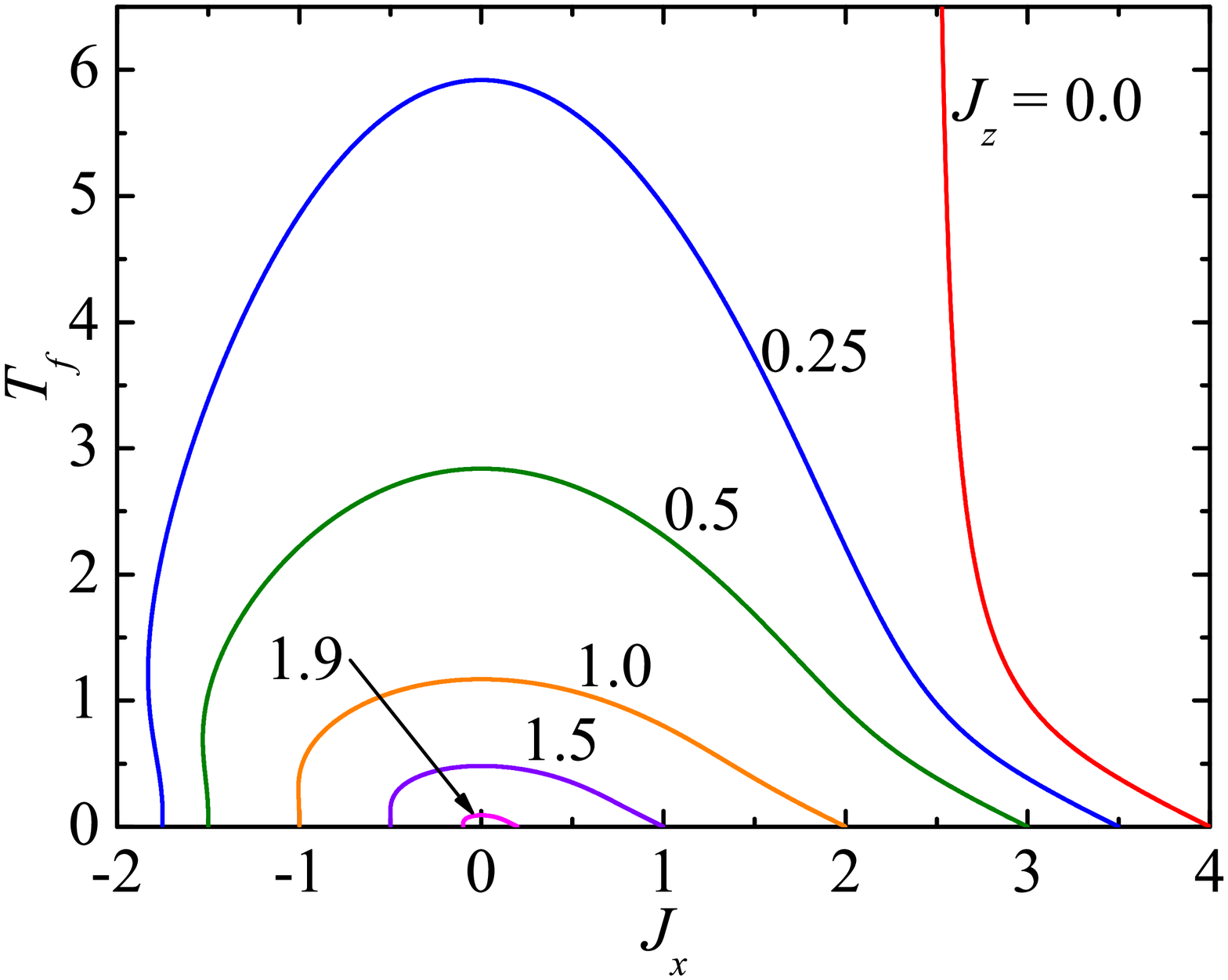}
\vspace{-0.5cm}
\caption{The frustration temperature $T_f$ as a function of the transverse component of the Heisenberg coupling $J_x$ for a few fixed values of its longitudinal component $J_z$.}
\label{fig4}
\end{figure}

\begin{figure}[!thb]
\centering
\includegraphics[scale=0.25]{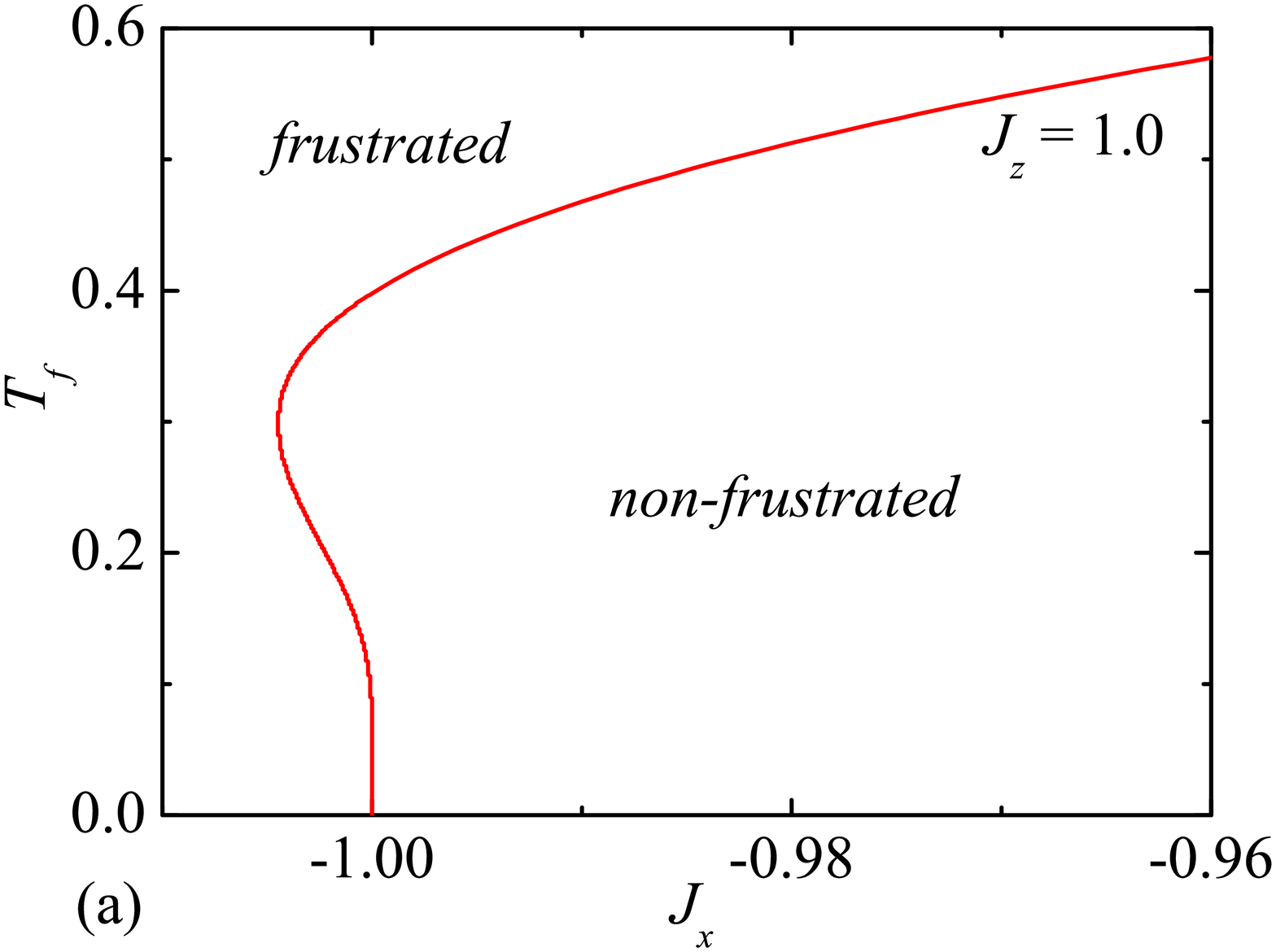}
\includegraphics[scale=0.25]{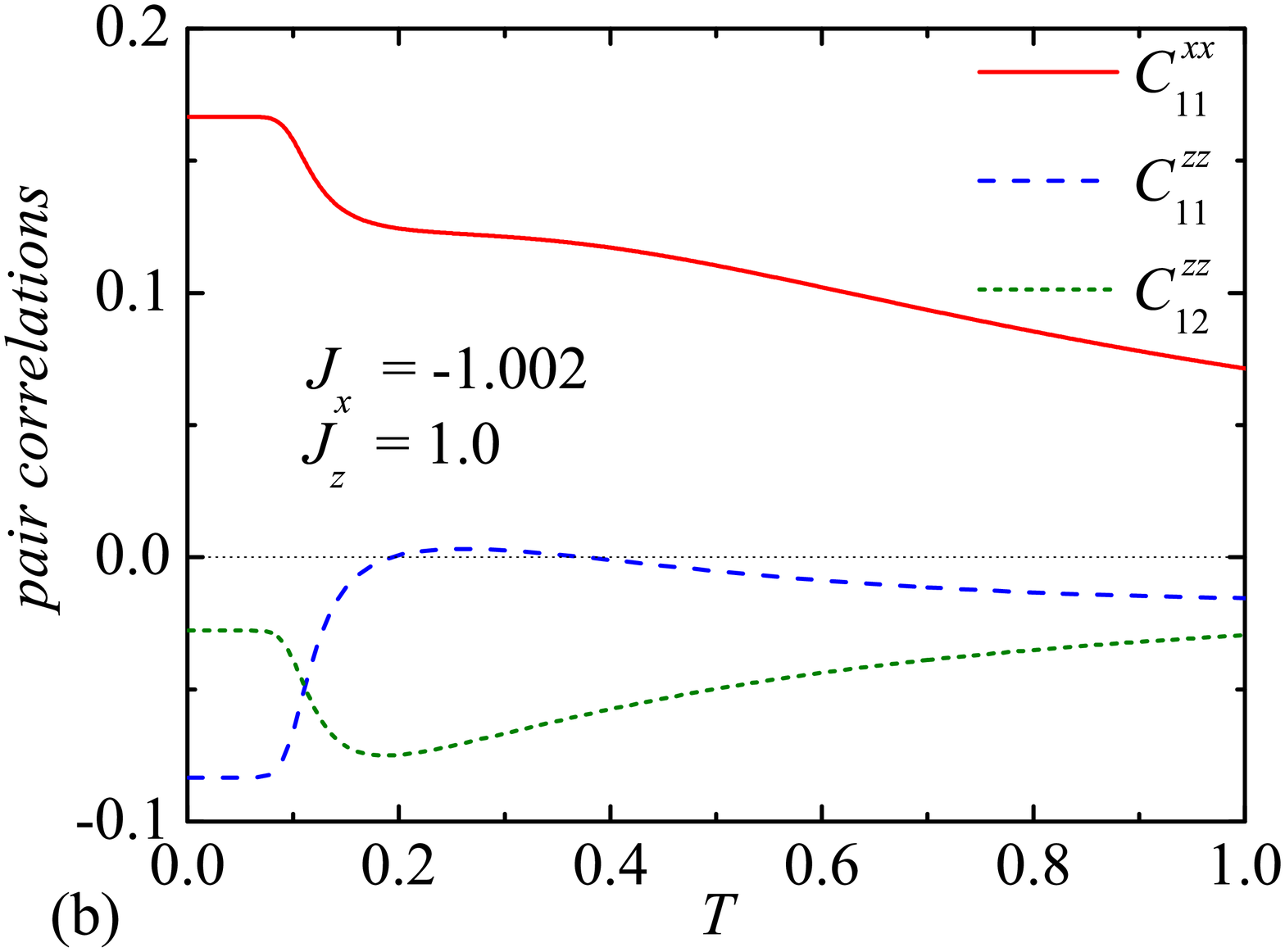}
\caption{(a) A reentrant behavior of the frustrated temperature $T_f$ in a vicinity of the ground-state phase boundary CAF-SQT for the particular case $J_z=1.0$; (b) Temperature dependences of the correlation functions for the parameter set $J_x= -1.002$ and $J_z = 1.0$ serving in evidence of the reentrant behavior (a thin dotted line at zero is guide for eyes only).}
\label{fig5}
\end{figure}

With this in mind, it might be quite interesting to investigate how the thermally activated spin frustration above the unfrustrated CAF ground state depends on a relative strength of the Heisenberg intra-triangle interaction. For this purpose, we have depicted in Fig. \ref{fig4} typical dependences of the frustration temperature $T_f$ on the transverse component $J_x$ of the Heisenberg coupling for a few fixed values of its longitudinal component $J_z$. It is worthwhile to recall that $z$-components of the spins from the same Heisenberg triangle are ferromagnetically (antiferromagnetically) correlated below (above) the frustration temperature $T_f$. In this regard, the spin-1/2 Ising-Heisenberg three-leg tube is free of frustration inside of the region bounded from above by the line of frustration temperatures, while it becomes frustrated outside of this region. It is evident from Fig. \ref{fig4} that the unfrustrated region gradually diminishes upon increasing  of the longitudinal component of the Heisenberg coupling until it completely disappears for any $J_z \geq 2$. This is agreement with absence of the unfrustrated CAF phase in the parameter region $J_z \geq 2$. It should be pointed out, moreover, that the upper- and lower-edge boundaries of the unfrustrated region exactly coincide at zero temperature with the ground-state boundaries CAF-DCA and CAF-SQT, respectively (c.f. Fig. \ref{fig4} with Fig. \ref{fig2}). 

Another interesting point to observe from Fig. \ref{fig4} is that the frustration temperature exhibits a notable reentrant behavior near its lower edge closely connected to the ground-state boundary between the CAF and SQT phases. To clarify this issue in a more detail, we have plotted in Fig. \ref{fig5} typical dependence of the frustration temperature in a close neighborhood of its lower-edge boundary [Fig. \ref{fig5}(a)] along with the corresponding thermal variations of the correlation functions [Fig. \ref{fig5}(b)]. If the transverse component of the Heisenberg interaction is selected sufficiently close but slightly below the ground-state boundary CAF-SQT, then, the longitudinal correlation function between the spins from the same Heisenberg triangle actually shows a weak ferromagnetic correlation within a relatively narrow range of moderate temperatures and antiferromagnetic correlation out of this temperature range. 

\subsection{Thermal entanglement}

\begin{figure}[!thb]
\centering
\includegraphics[scale=0.25]{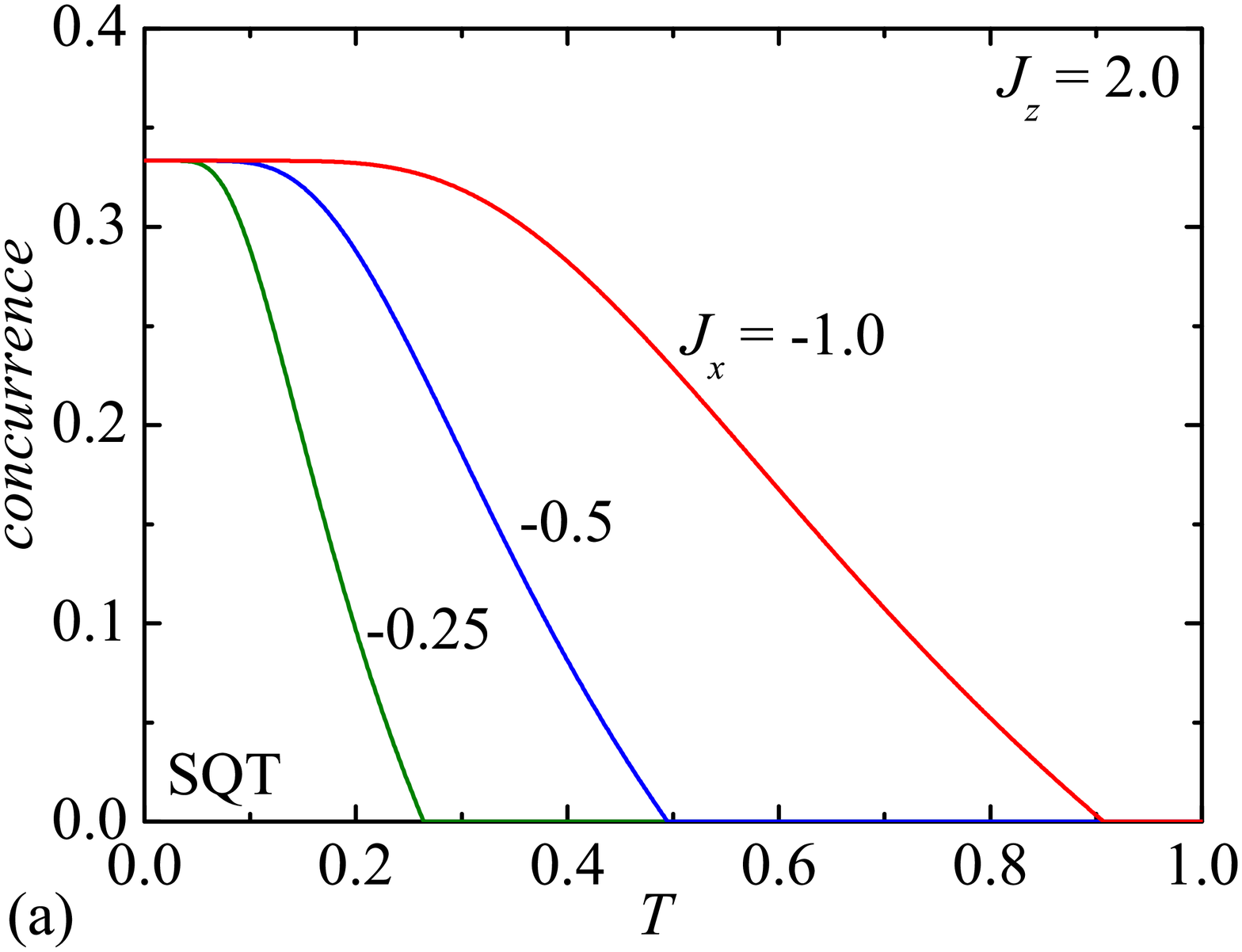}
\includegraphics[scale=0.25]{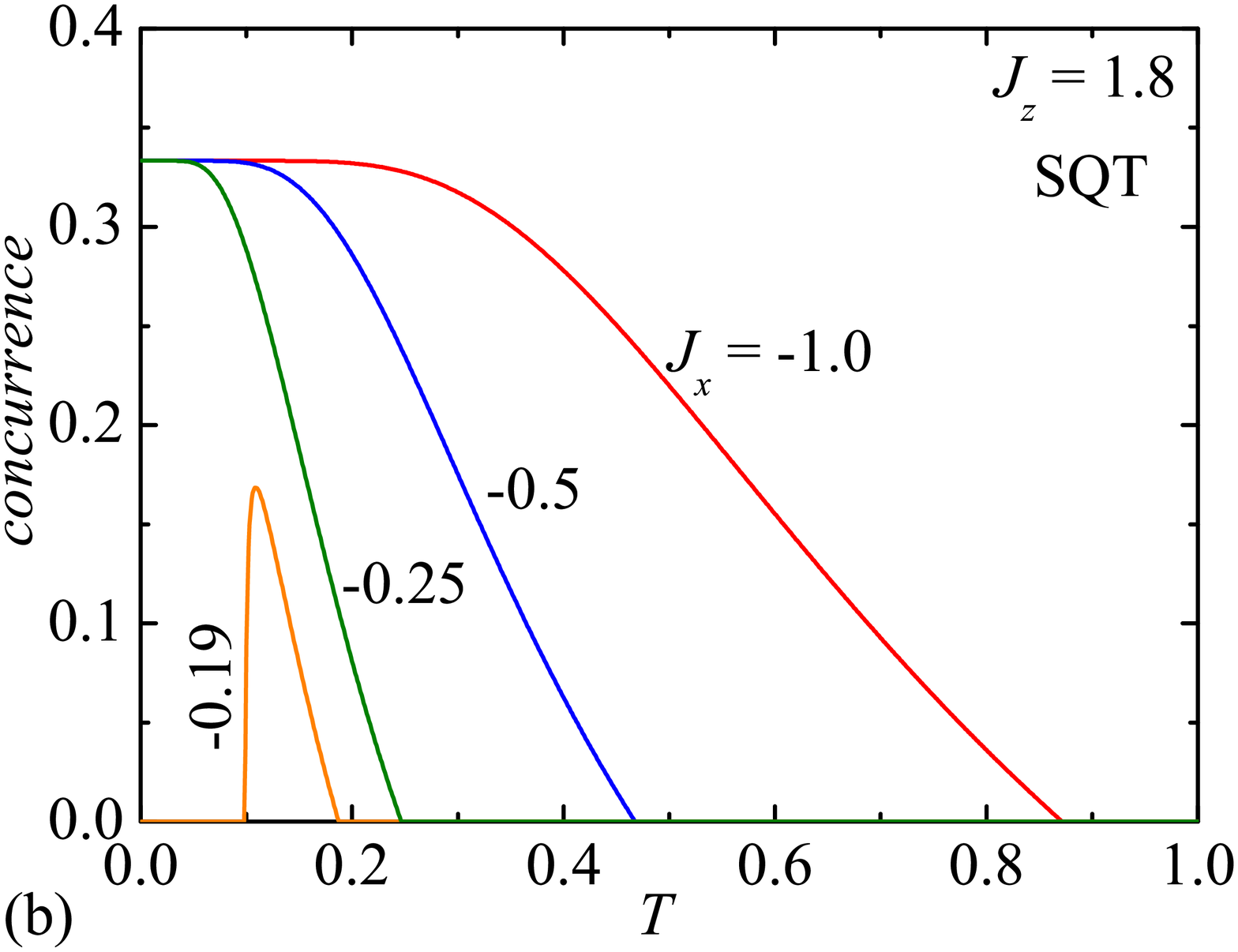}
\vspace{-0.5cm}
\caption{Thermal variations of the concurrence for several values of the transverse component of the Heisenberg intra-triangle coupling $J_x$ and two different values of its longitudinal component: (a) $J_z=2.0$; (b) $J_z=1.8$.}
\label{fig6}
\end{figure}

The concurrence, as calculated from Eq. (\ref{conc}), represents a feasible measure of bipartite quantum entanglement at zero as well as non-zero temperatures. Although the absence of quantum correlations in the CAF ground state could be anticipated on the grounds of the fully classical character of this phase, it is somewhat more surprising that the concurrence equals zero also within the DCA ground state. According to this, the SQT phase is the only ground state for which the calculated concurrence $\mathrm{C} = 1/3$ at zero temperature indicates the substantial but not full quantum entanglement. To clarify the effect of temperature upon the bipartite entanglement, we have plotted in Fig. \ref{fig6} the concurrence against temperature for two different values of the longitudinal component $J_z$ of the Heisenberg intra-triangle coupling and several values of its transverse component $J_x$. In agreement with the general expectations, thermal fluctuations gradually destroy the quantum entanglement, i.e. the concurrence generally decreases upon increasing temperature until it finally disappears above a certain temperature referred to as the threshold temperature $T_t$. Apart from this standard dependence, one may also found a peculiar reentrant behavior of the concurrence, which is illustrated in Fig. \ref{fig6}(b) on the particular example with $J_z=1.8$ and $J_x=-0.19$. Under this circumstance, the concurrence evolves from zero just at a certain lower threshold temperature, then it shows a peculiar thermally-induced increase followed by a successive thermally-induced decrease until it completely vanishes at an upper threshold temperature. 

\begin{figure}[!thb]
\centering
\includegraphics[scale=0.25]{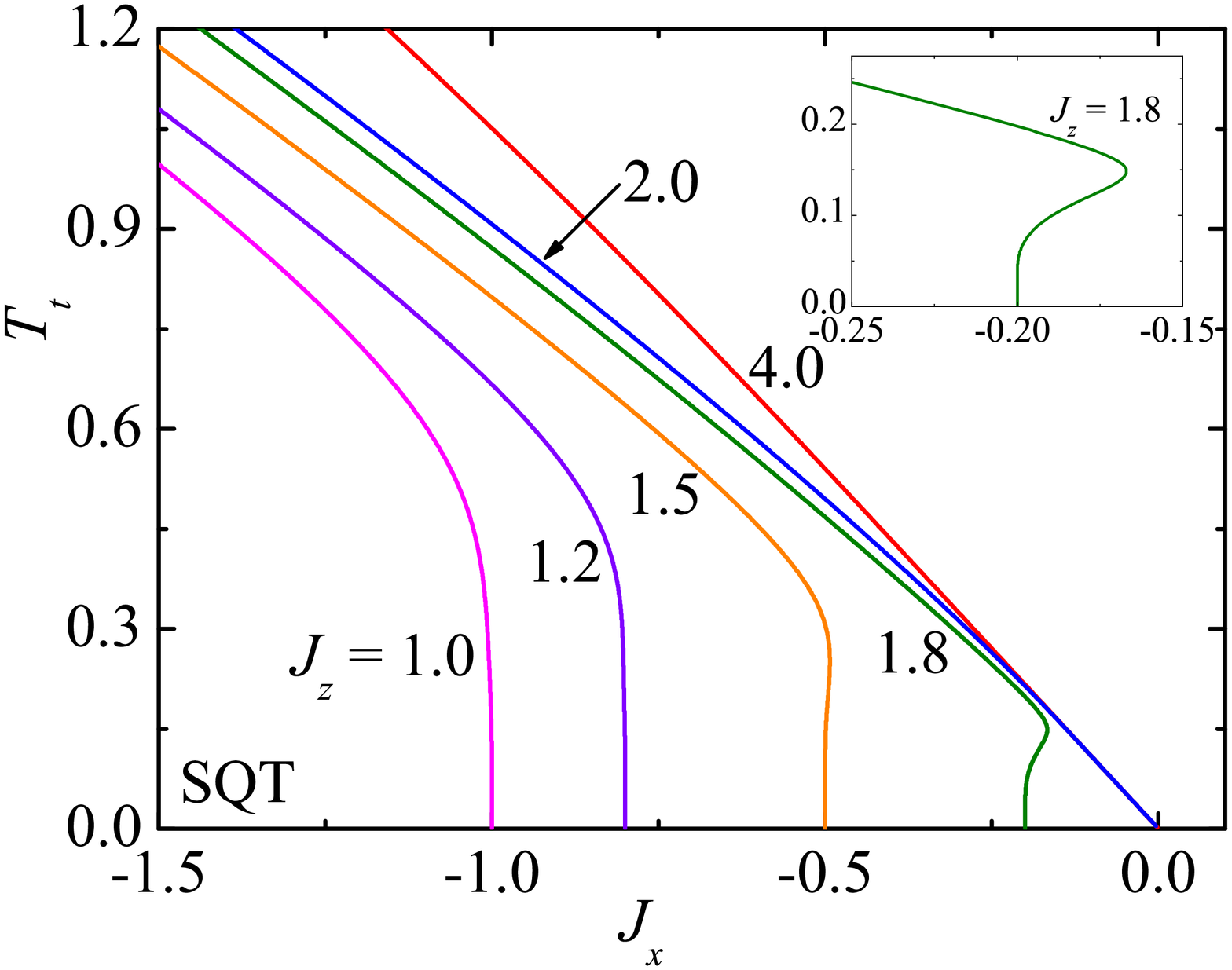}
\vspace{-0.5cm}
\caption{The threshold temperature $T_t$ as a function of the transverse component of the Heisenberg coupling $J_x$ for a few fixed values of its longitudinal component $J_z$. The inset shows a detail from the reentrant region for the particular case $J_x = 1.8$.}
\label{fig7}
\end{figure}

To gain an overall insight into the entangled part of the parameter region, we have depicted in Fig. \ref{fig7} the threshold temperature as a function of the transverse component $J_x$ of the Heisenberg intra-triangle coupling for several fixed values of its longitudinal component $J_z$. The spin-1/2 Ising-Heisenberg tube is entangled inside of the parameter region bounded from above by displayed lines of the threshold temperature, where the concurrence as a measure of the thermal entanglement is non-zero. If the longitudinal component of the antiferromagnetic Heisenberg coupling is sufficiently strong  $J_z \geq 2$, then, the threshold temperature monotonically decreases with increasing its transverse component $J_x$ until it tends to zero at $J_x=0$. On the other hand, the dependence of the threshold temperature terminates for $J_z<2$ at the ground-state phase boundary between the SQT and CAF phases at $J_x=J_z-2$. Moreover, it can be observed from Fig. \ref{fig7} that the threshold temperature shows the most striking dependence with a pronounced reentrant region when the longitudinal component of the Heisenberg intra-triangle interaction is close enough but slightly below 
$J_z \lessapprox 2$. 

\subsection{Frustration vs. entanglement}

\begin{figure}[!thb]
\centering
\includegraphics[scale=0.25]{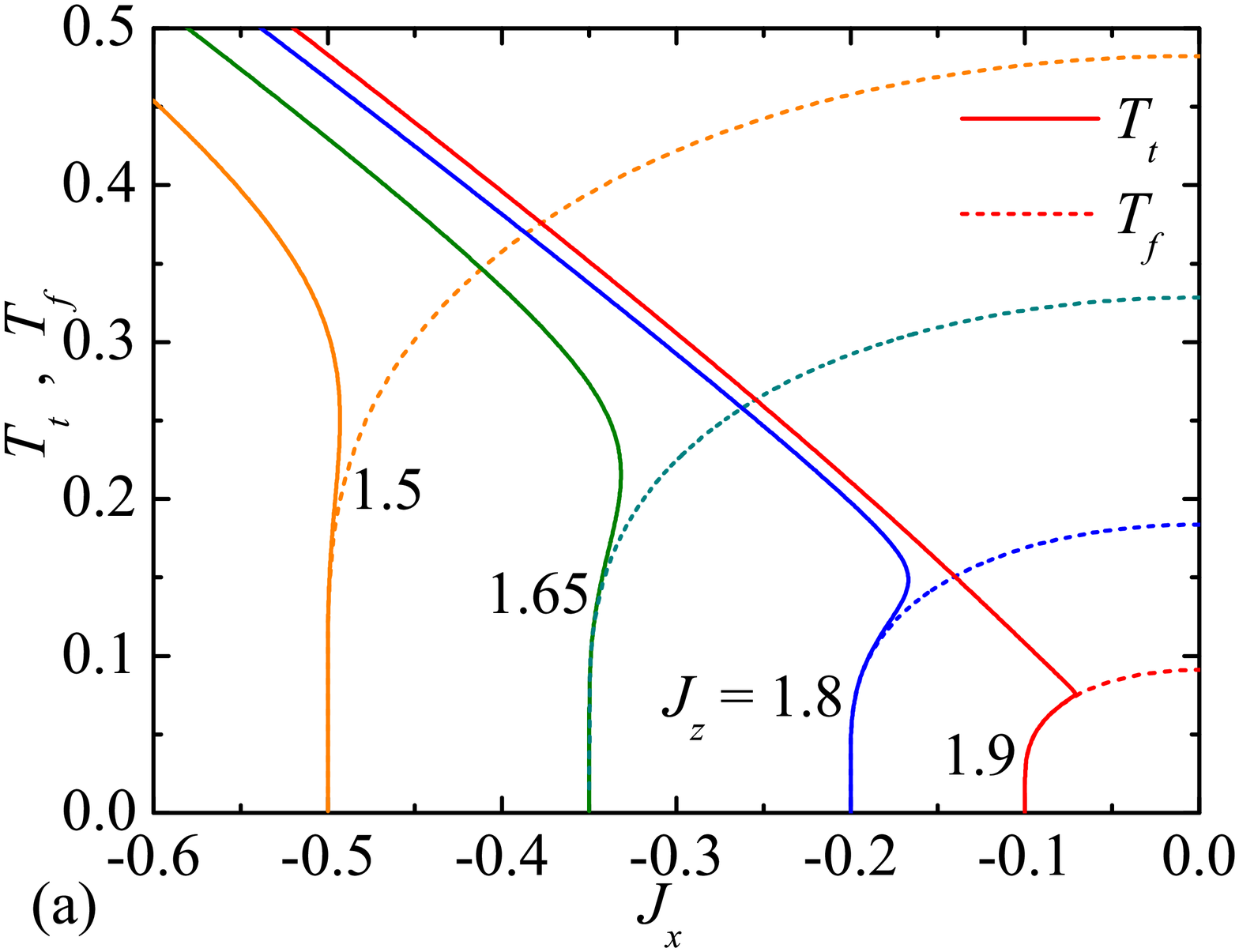}
\includegraphics[scale=0.25]{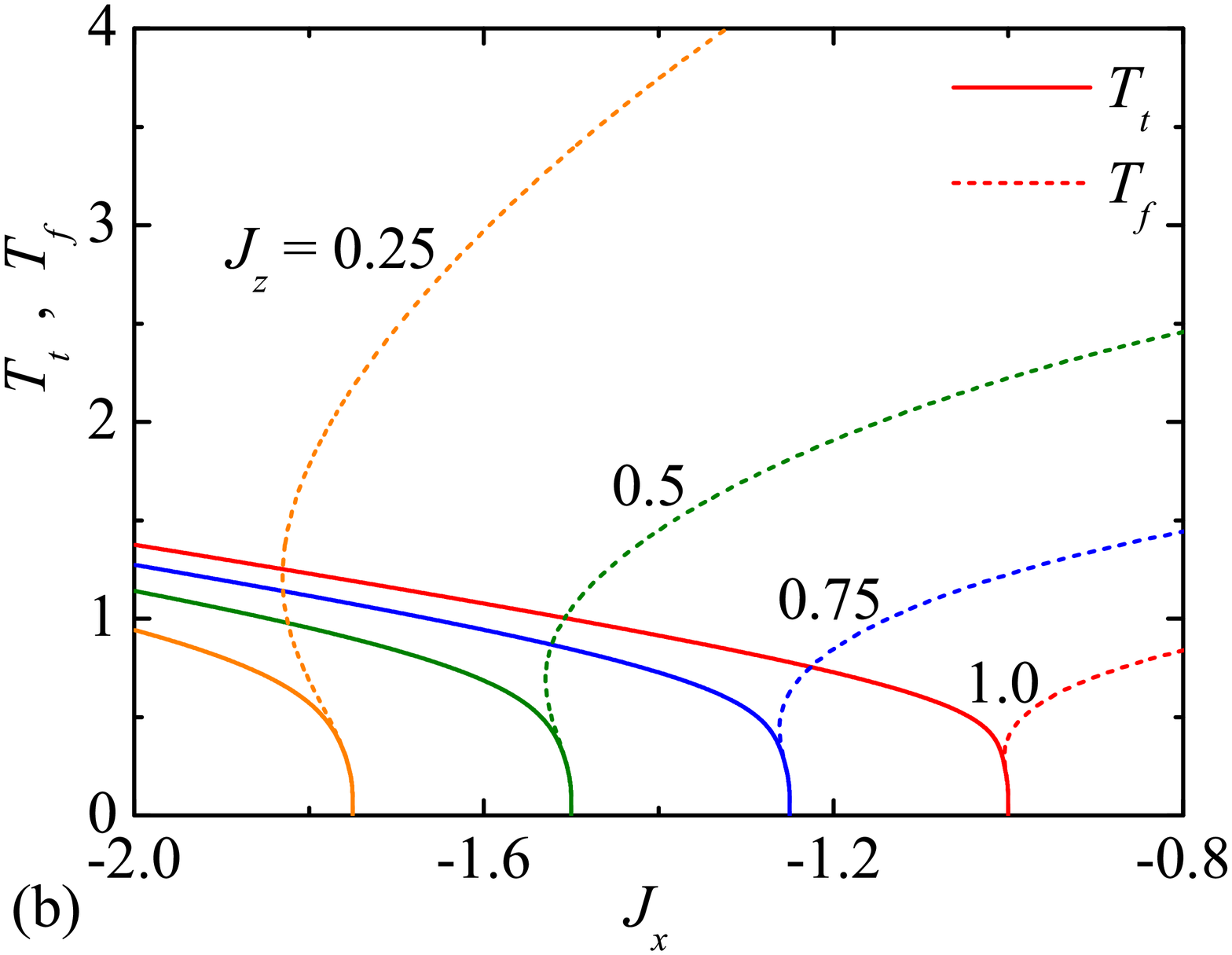}
\vspace{-0.5cm}
\caption{The dependence of threshold (solid lines) and frustration (broken lines) temperatures on the transverse component of the Heisenberg intra-triangle coupling $J_x$ for several fixed values of its longitudinal component $J_z$. The panel (a) shows reentrant behavior of the threshold temperature and the panel (b) reentrant behavior of the frustration temperature.}
\label{fig8}
\end{figure}

At this stage, it might be of particular interest to investigate a mutual interplay between the thermally activated spin frustration and entanglement, which do not bear at first sight any direct relation. To this end, we have plotted in Fig. \ref{fig8} the threshold and frustration temperature against the transverse component of the Heisenberg intra-triangle coupling $J_x$ for several fixed values of its longitudinal component $J_z$. It is quite apparent from this comparison that the threshold and frustration temperatures coincide at low enough temperatures, because they both converge to the identical zero-temperature asymptotic limit though they show completely different behavior at higher temperatures. It can be also understood from Fig. \ref{fig8} that the thermal entanglement occurs just outside of the parameter region bounded by the line of frustration temperatures, 
which means that the spin frustration is in the spin-1/2 Ising-Heisenberg tube indispensable for a presence of the thermal entanglement. Another interesting point is that the reentrance in the threshold temperatures gives rise to the thermal entanglement above the unfrustrated parameter space for $J_z \lessapprox 2$ [see Fig. \ref{fig8}(a)], while the reentrance in the frustration temperatures makes possible to detect the unfrustrated region above the entangled parameter space [see Fig. \ref{fig8}(b)]. Both types of reentrances are apparently antagonistic and cannot emerge simultaneously. 

\subsection{Quantum non-locality}

\begin{figure}[!thb]
\centering
\includegraphics[scale=0.25]{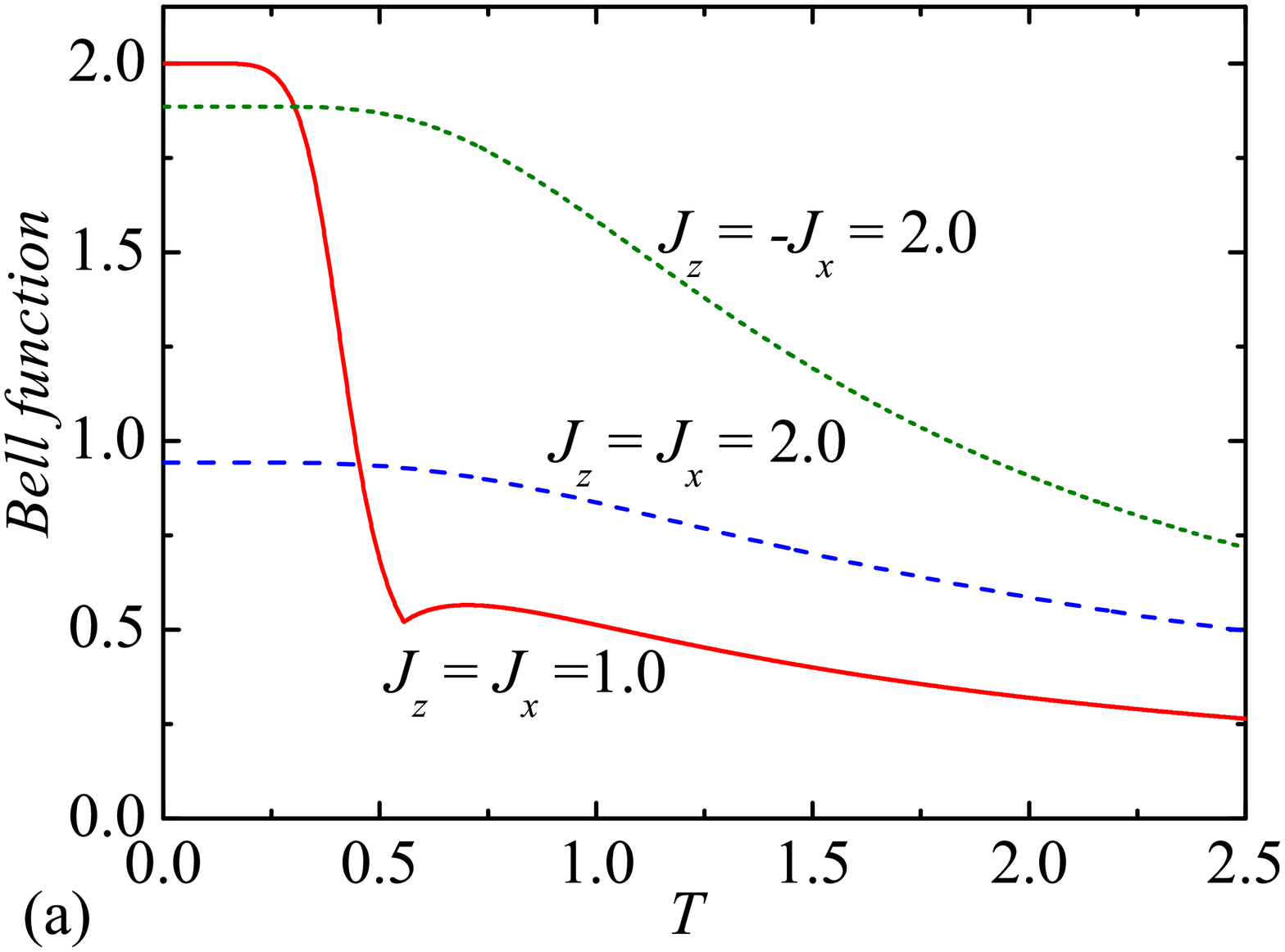}
\includegraphics[scale=0.25]{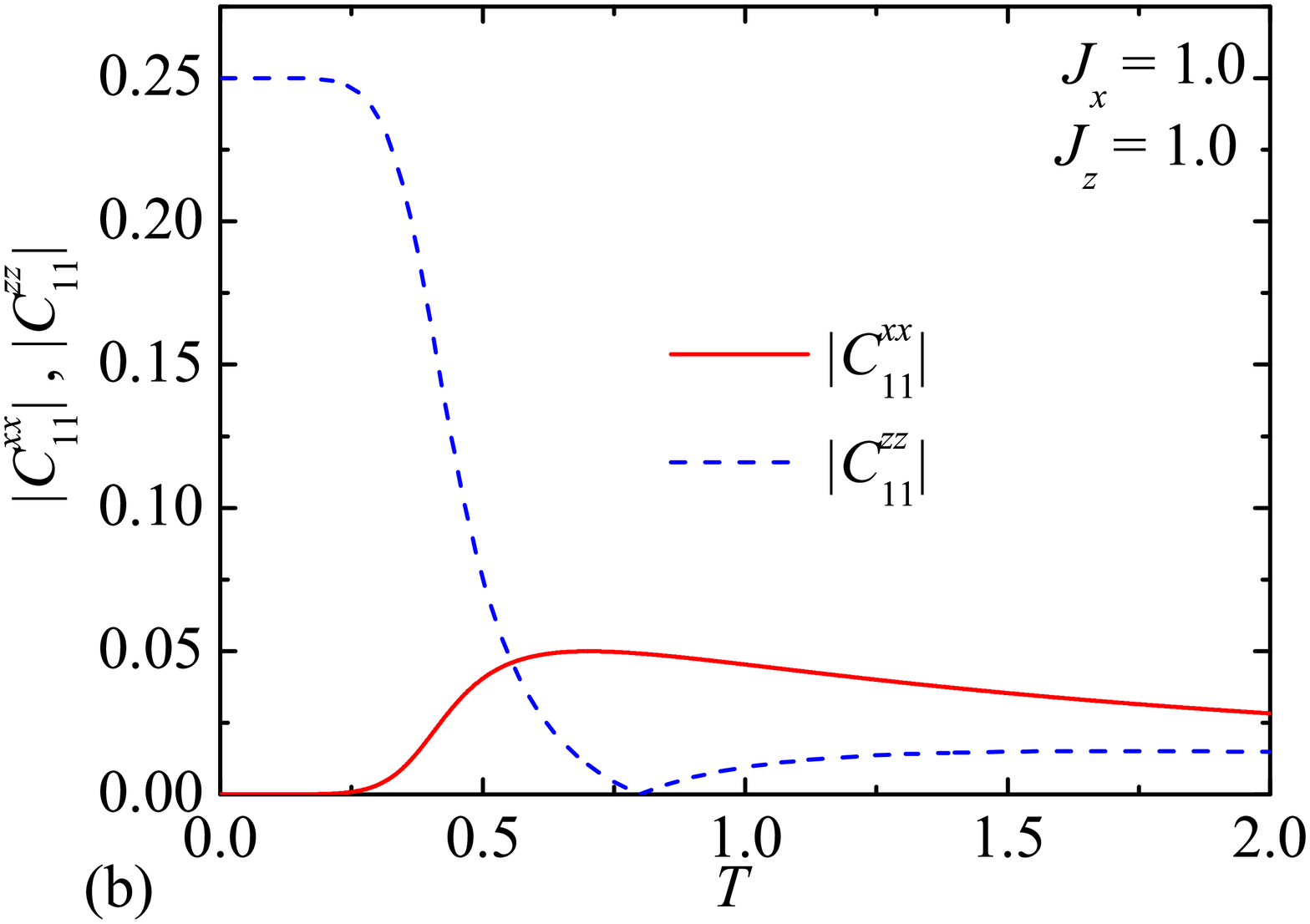}
\vspace{-0.5cm}\caption{(a) The Bell function versus temperature for three different Heisenberg intra-triangle interactions corresponding to three available ground states: $J_z=J_x=1.0$ (CAF phase), $J_z=-J_x=2.0$ (SQT phase), $J_z=J_x=2.0$ (DCA phase); (b) Thermal variations of absolute values of the longitudinal and transverse pair correlation function between the spins from the same Heisenberg triangle for the parameter set $J_z=J_x=1.0$ (CAF phase).}
\label{fig9}
\end{figure}

Next, it could be quite interesting to answer the question whether or not the spin-1/2 Ising-Heisenberg tube may violate the Bell inequality, because the entanglement and non-locality capture closely related but independent features of quantum correlations. A comprehensive analysis reveals that all three available ground state do not violate the Bell inequality, since the calculated value of the Bell function never exceeds the largest value $\mathrm{B} = 2$ allowed for classical correlations. To support this statement, we have depicted in Fig. \ref{fig9}(a) typical temperature variations of the Bell function for three different sets of the Heisenberg intra-triangle interaction, which drive the investigated model to the CAF, SQT and DCA ground states, respectively. Altogether, it could be concluded that quantum correlations are in the spin-1/2 Ising-Heisenberg tube strictly local in spite of the fact that the thermal entanglement is evidently present within the SQT ground state. The cuspate dip of the Bell function at the temperature $T \approx 0.55$ of the particular case with the CAF ground states thus represents the most striking feature of the displayed dependences. A presence of this kind of singularity can be attributed to a crossing of absolute values of the transverse and longitudinal pair correlation function between the spins from the same Heisenberg triangle, which indicates according to Eq. (\ref{bell}) two different analytic prescriptions below and above a relevant crossing point.

\subsection{Specific heat and entropy}

\begin{figure}[!thb]
\centering
\includegraphics[scale=0.25]{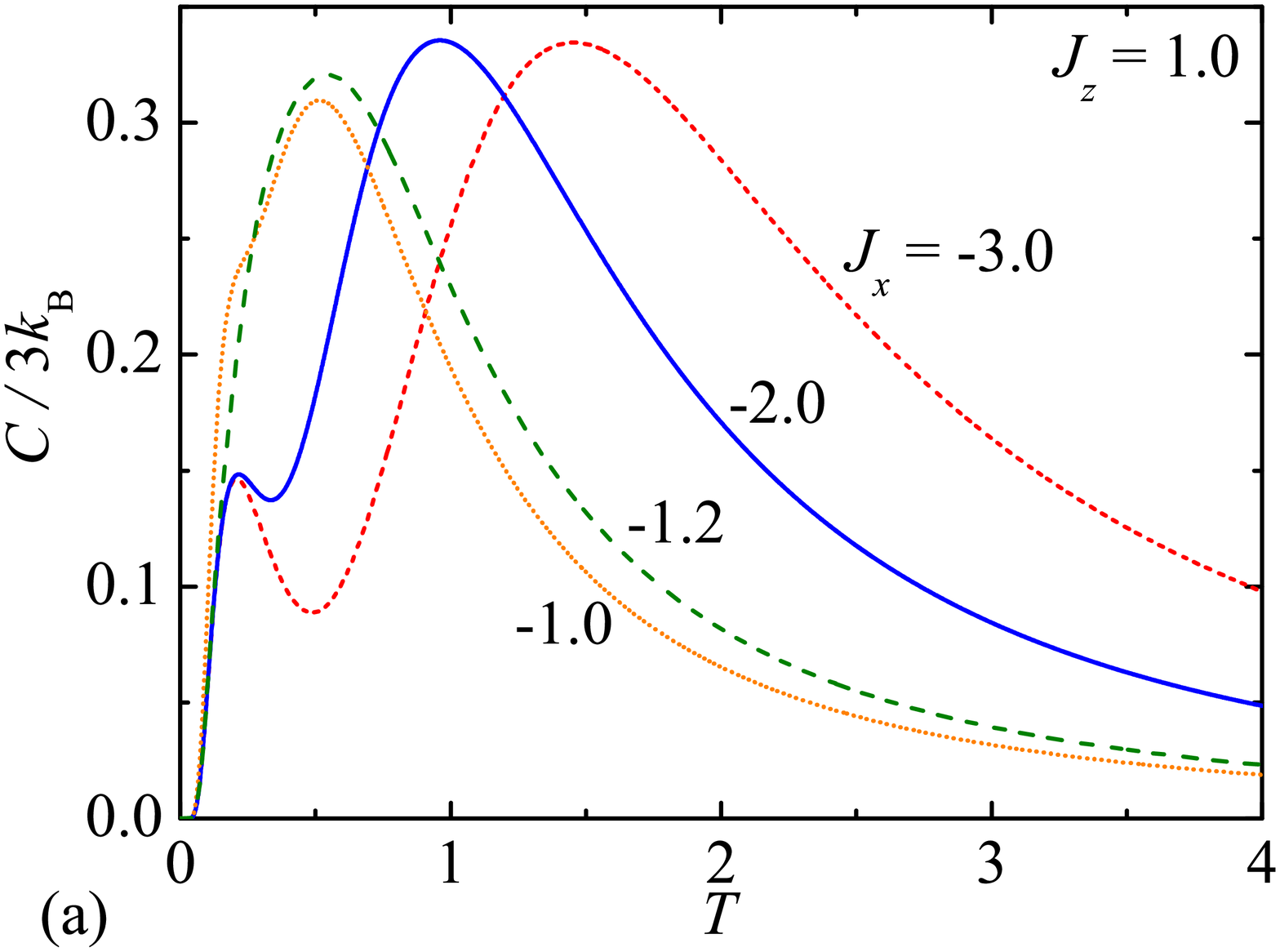}
\includegraphics[scale=0.25]{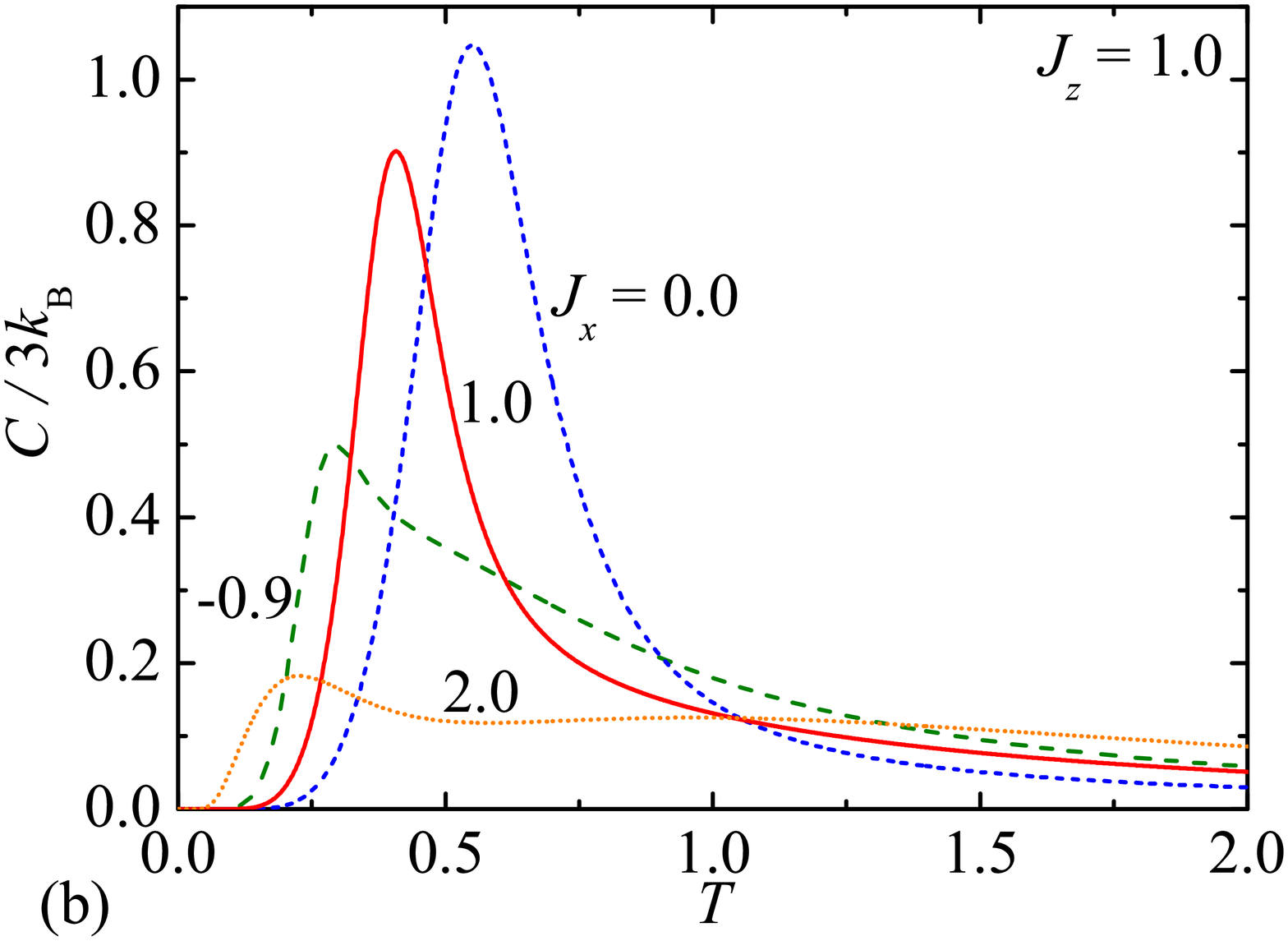}
\includegraphics[scale=0.25]{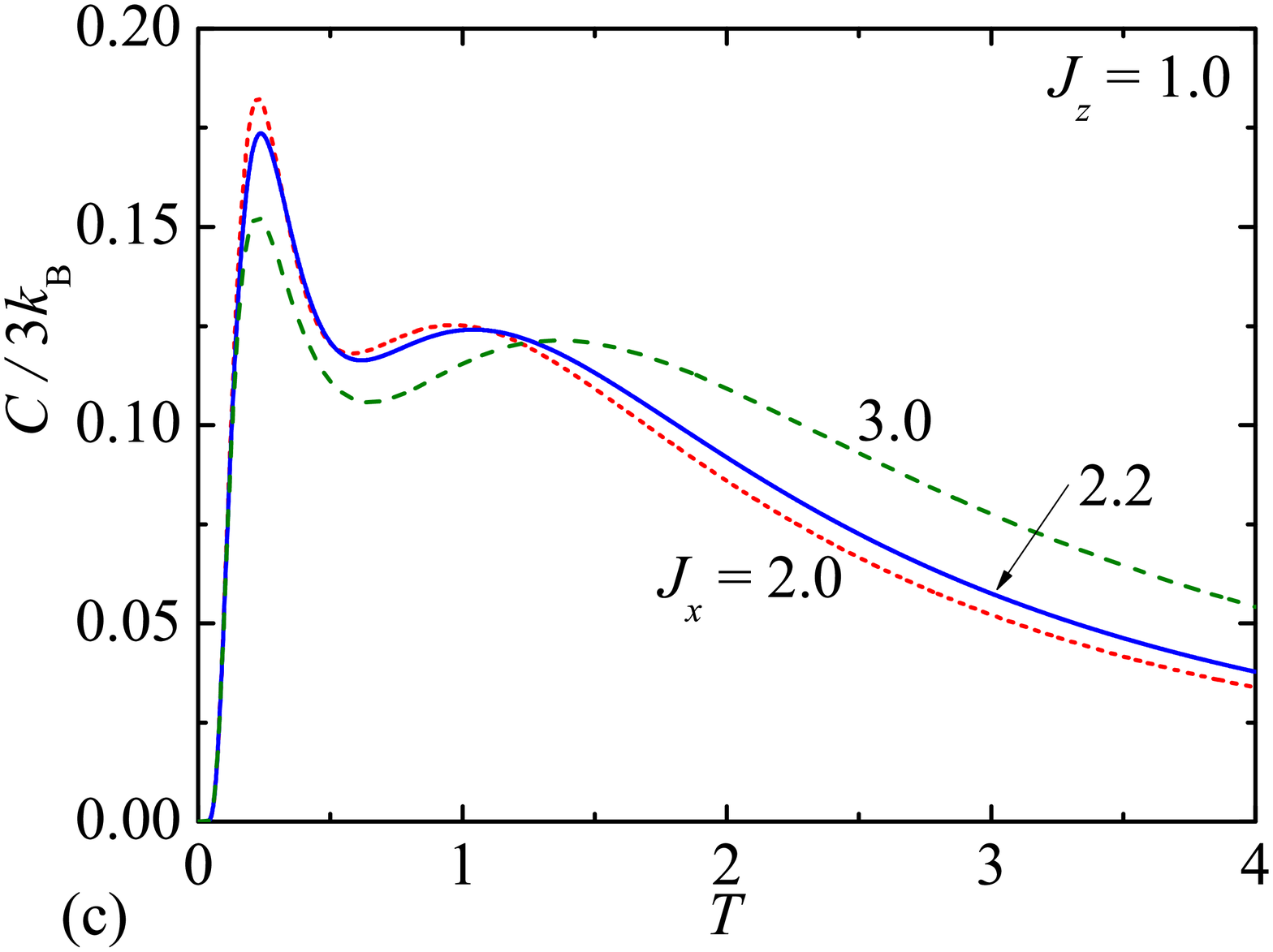}
\vspace{-0.5cm}
\caption{Temperature variations of the specific heat (per one spin) for the fixed value of the longitudinal component of the Heisenberg coupling $J_z=1$ and several values of its transverse component $J_x$. The selected coupling constants $J_x$ are consistent with the following ground states: (a) SQT phase; (b) CAF phase; (c) DCA phase.}
\label{fig10}
\end{figure}

The substantial thermal variations of the correlation functions near the ground-state phase boundaries may manifest themselves also in unusual temperature dependences of basic thermodynamic quantities, so let us explore in the following typical thermal variations of the zero-field specific heat.
It can be seen from Fig. \ref{fig10}(a) that the specific heat can exhibit a peculiar double-peak temperature dependence when the SQT phase constitutes the ground state, whereas the low-temperature peak predominantly comes from thermally-induced breakdown of the longitudinal correlation between the spins from the neighboring triangles. Note furthermore that the low-temperature peak gradually merges with the round high-temperature maximum, which shifts to lower temperatures when the spin system approaches the ground-state phase boundary between the SQT and CAF phases (at $J_x=-1$ when $J_z=1$ is fixed).

Contrary to this, the specific heat shows a more common temperature dependence with a single maximum in a majority of the parameter space, which corresponds to the CAF ground state [Fig. \ref{fig10}(b)]. The only notable exception from this rule is when the Heisenberg intra-triangle coupling drives the spin system sufficiently close to the ground-state phase boundary with the DCA phase at $J_x=2.0$ assuming the fixed value of $J_z=1.0$ (see the subsequent paragraph). Last but not least, one recovers the more striking double-peak temperature dependence of the specific heat on assumption that the DCA phase constitutes the ground state [Fig. \ref{fig10}(c)]. Under this condition, the low-temperature peak predominantly comes from the thermally-assisted breakdown of the longitudinal correlation between the spins from the neighboring triangles. 

\begin{figure}[!thb]
\centering
\includegraphics[scale=0.25]{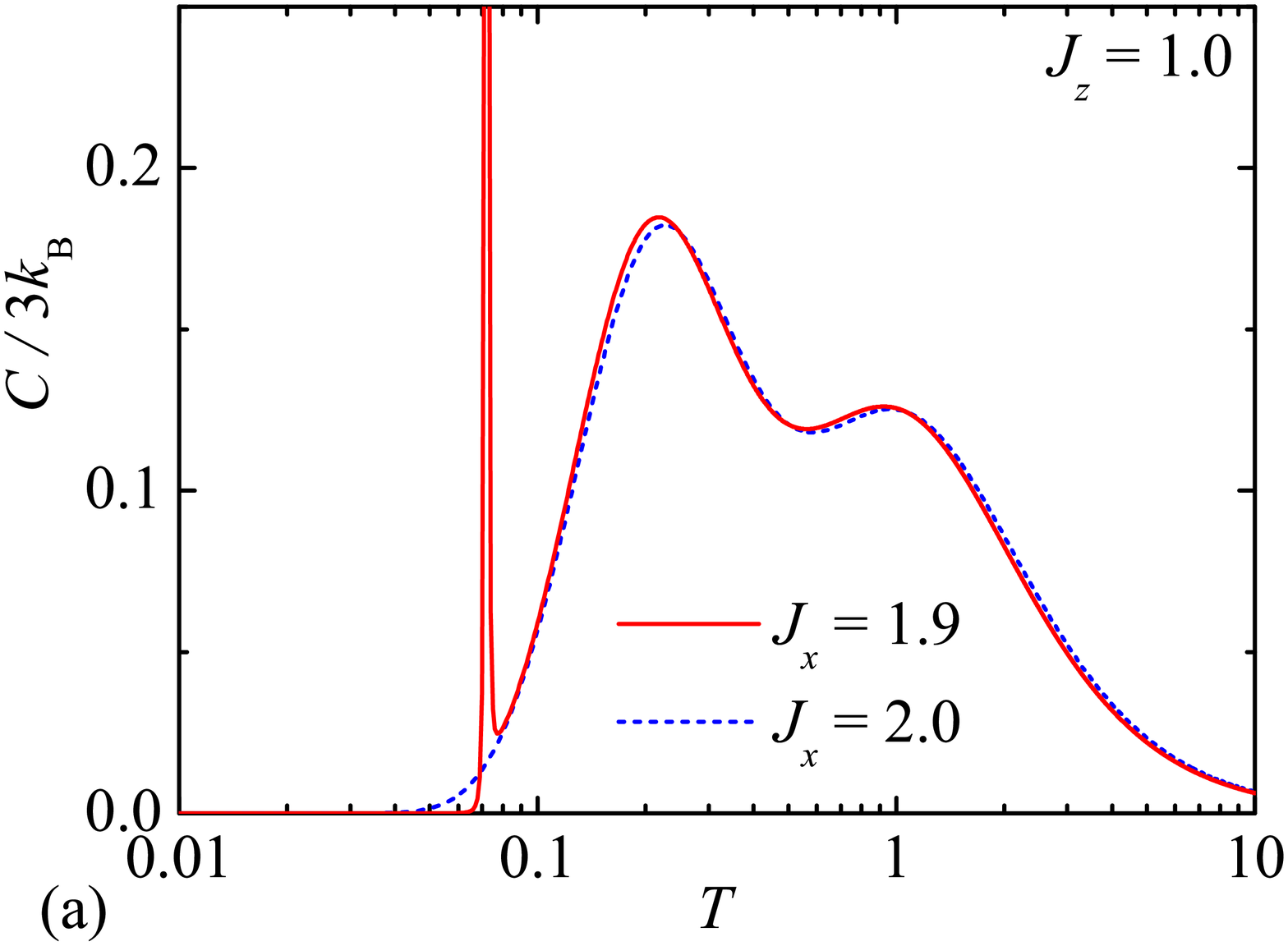}
\includegraphics[scale=0.25]{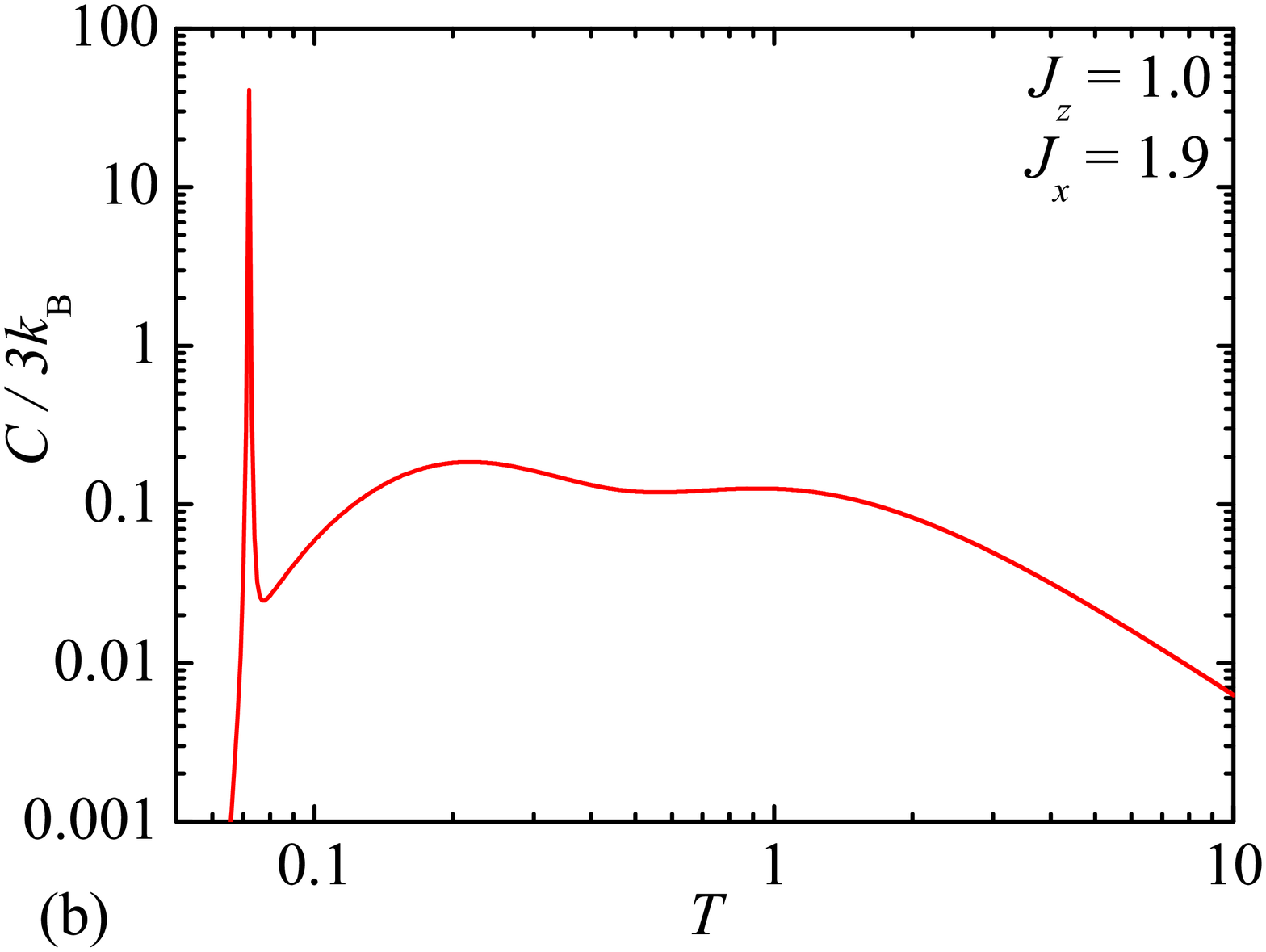}
\includegraphics[scale=0.25]{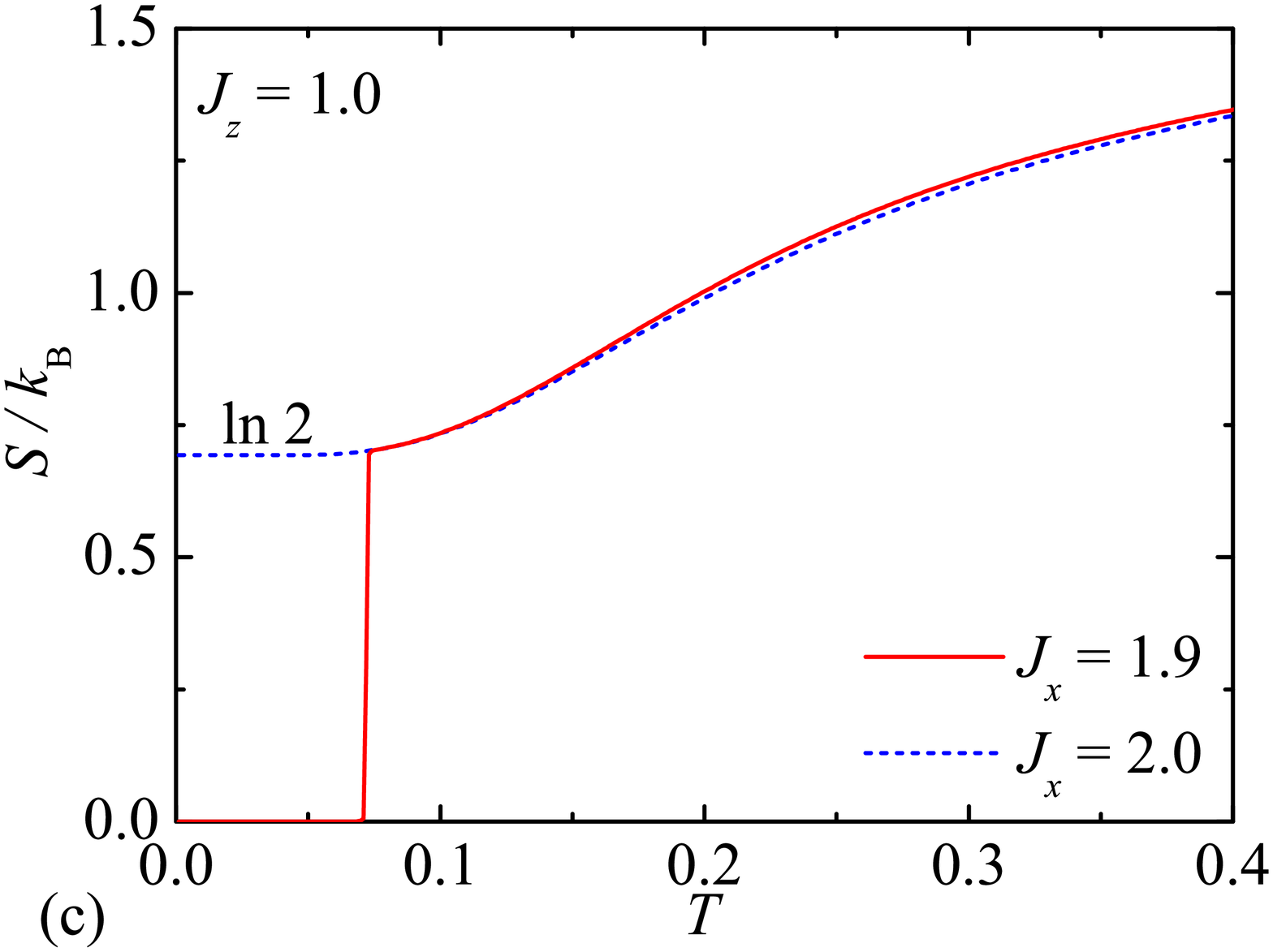}
\vspace{-0.5cm}
\caption{(a) The semi-logarithmic plot for the temperature dependence of the specific heat (per one spin) exactly at the ground-state phase boundary CAF-DCA ($J_x=2.0; J_z=1.0$) and just below it ($J_x=1.9; J_z=1.0$); (b) The low-temperature peak of the specific heat for the case with  $J_z=1.0$ and $J_x=1.9$ in a log-log scale; (c) The temperature dependence of the entropy (per one triangle) exactly at the ground-state phase boundary CAF-DCA ($J_x=2.0; J_z=1.0$) and just below it ($J_x=1.9; J_z=1.0$).}
\label{fig11}
\end{figure}

Let us turn back to the most spectacular temperature dependence of the specific heat, which involves three separate peaks as displayed in Fig. \ref{fig11}(a)-(b). The triple-peak thermal dependence of the zero-field specific heat can be found when the Heisenberg intra-triangle coupling drives the spin-1/2 Ising-Heisenberg tube towards the CAF ground state but still keeps it in a close vicinity of the phase boundary with the DCA phase (at $J_x=2.0$ for $J_z=1.0$). While thermal excitations of physically different origin are responsible for an existence of the high-temperature maximum at $T \approx 1.0$, the round maximum at moderate temperatures $T \approx 0.25$ relates to a gradual decline of the longitudinal and transverse correlations between the spins from the same triangle. The most surprising is of course a presence of the sizable low-temperature peak, which could be at first sight easily confused with a temperature-driven first-order phase transition. The sharp and very narrow low-temperature peak, which emerges around the temperature $T \approx 0.072$ by considering the particular case with $J_x=1.9$ and $J_z=1.0$, can be ascribed to massive thermal excitations from the two-fold degenerate CAF ground state to the macroscopically degenerate DCA excited state. As a matter of fact, the locus of the sharp low-temperature peak is in a good concordance with the formula
\begin{eqnarray}
T_p = \frac{4 - 2J_z - J_x}{\ln 4},
\label{peak}
\end{eqnarray}
which follows from a direct comparison of the Helmholtz free energies of the CAF and DCA phases provided that thermal variations of the internal energy and entropy are simply neglected. Thus, it could be concluded that the sharp low-temperature peak of the specific heat appears due to a high entropy gain, which originates from the chiral degrees of freedom of the DCA phase lying in energy just slightly above the doubly degenerate CAF ground state. To support this statement, we have plotted in Fig. \ref{fig11}(c) the relevant thermal variations of the entropy, which provides a convincing evidence for an abrupt but still continuous change of the entropy from almost zero to $\ln 2$ associated with the vigorous thermal excitations from the CAF phase to the DCA phase. The abrupt entropy change can be detected at the temperature, which is in accordance with the position of the sharp low-temperature peak of the specific heat given by the formula (\ref{peak}).

\section{Conclusion}
\label{conclusion}

In the present work, we have exactly solved the spin-1/2 Ising-Heisenberg three-leg tube by taking advantage of the local conservation of the total spin
on each Heisenberg spin triangle and the classical transfer-matrix method. The elaborated rigorous procedure has enabled us to derive exact results for the ground-state phase diagram, basic thermodynamic quantities and several pair correlation functions, which were subsequently employed for a calculation of the concurrence and Bell function. The latter two quantities were used in order to quantify thermal entanglement and non-locality, which are related to quantum correlations between two spins coupled by the Heisenberg intra-triangle interaction. While none of three available ground states violates the Bell inequality, the SQT phase with a regular alternation of the symmetric quantum superposition of up-up-down and down-down-up on odd and even triangles (or vice versa) does exhibit the thermal entanglement.  

It has been demonstrated that the SQT and DCA ground states are naturally frustrated unlike the unfrustrated CAF ground state, above which the so-called thermally activated spin frustration can develop provided that the antiferromagnetic intra-triangle interaction $J_z>0$ is considered. We have rigorously calculated the frustration temperature delimiting the unfrustrated region from the frustrated one, which was subsequently compared with the threshold temperature of a disappearance of the thermal entanglement. It has been verified that the frustration and threshold temperatures coincide at sufficiently low temperatures though they exhibit a very different at higher temperatures. Moreover, it turns out that the spin-1/2 Ising-Heisenberg three-leg tube is thermally entangled just in the frustrated region, which implies that the frustration represents indispensable ground for a presence of the thermal entanglement in this spin system. Note furthermore that the lines of threshold and frustration temperatures may display a reentrant phenomenon though both reentrances are antagonistic and cannot appear simultaneously. Hence, the famous dictum that quantum correlations are gradually suppressed through thermal fluctuations is not of general validity, because thermal fluctuations can alternatively act against classical spin arrangements and thus leaving more space to an emergence of the thermal entanglement above a classical ground state.

The most interesting finding stemming from our study certainly represents an extraordinary diversity of temperature dependences of the zero-field specific heat, which may show up to three separate local maxima. The most remarkable temperature variations of the specific heat involve a sharp low-temperature peak extended in a very narrow temperature range, which is quite reminiscent of a temperature-driven first-order phase transition. However, it has been convincingly evidenced that this anomalous peak relates to massive thermal excitations from the doubly degenerate CAF phase to the macroscopically degenerate DCA phase with two chiral degrees of freedom per each Heisenberg spin triangle. To the best of our knowledge, the spin-1/2 Ising-Heisenberg three-leg tube
is just the second example of the exactly solved model with such an intriguing feature in addition to a hybrid spin-electron double-tetrahedral chain \cite{gali15}. 

Finally, it should be also mentioned that the rigorous procedure elaborated in the present work can be straightforwardly adapted to account for the non-zero external magnetic field as well. Our future work will continue in this direction.

\end{document}